\documentclass[12pt]{article}
\pdfoutput=1

\newlength{\abstractwidth}
\usepackage{amsmath}
\usepackage{amsfonts}
\usepackage{amssymb}
\usepackage{latexsym}

\usepackage{epsf}
\usepackage{color}
\usepackage{graphicx}
\usepackage{tikz}
\usepackage{dsfont}

\usetikzlibrary{arrows,shapes,positioning}
\usetikzlibrary{decorations.markings}
\usepackage[rightcaption]{sidecap}
\tikzstyle arrowstyle=[scale=1]
\tikzstyle directed=[postaction={decorate,decoration={markings,
    mark=at position .65 with {\arrow[arrowstyle]{stealth}}}}]
\tikzstyle reverse directed=[postaction={decorate,decoration={markings,
    mark=at position .65 with {\arrowreversed[arrowstyle]{stealth};}}}]
\usetikzlibrary{positioning}

\usepackage{hyperref}
\definecolor{darkred}{rgb}{0.8,0.1,0.1}
\hypersetup{colorlinks=true, linkcolor=darkred, citecolor=blue, linktoc=page}

\makeatletter\def\l@subsubsection#1#2{}%
\makeatother

\renewcommand{\thefootnote}{\fnsymbol{footnote}}
\renewcommand{\thanks}[1]{\footnote{#1}}
\newcommand{\starttext}{
\setcounter{footnote}{0}
\renewcommand{\thefootnote}{\arabic{footnote}}}

\newcommand{\bea}{\begin{eqnarray}}
\newcommand{\eea}{\end{eqnarray}}
\newcommand{\be}{\begin{eqnarray}}
\newcommand{\ee}{\end{eqnarray}}
\newcommand{\bal}{\begin{align}}
\newcommand{\eal}{\end{align}}
\newcommand{\bma}{\begin{matrix}}
\newcommand{\ema}{\end{matrix}}

\def\Im{{\rm Im \,}}

\def\det{{\rm det \,}}

\usepackage{bbold}

\usepackage{verbatim} 

\usepackage{titling}
\usepackage{cite}

\numberwithin{equation}{section} 

\begin{document}
\starttext
\setcounter{footnote}{0}

\title{{ Warped $AdS_{2}$ and $SU(1,1|4)$ symmetry in Type IIB}}
\author{{\bf David Corbino} \\ 
{\sl  Mani L. Bhaumik Institute for Theoretical Physics}\\
{\sl Department of Physics and Astronomy} \\
{\sl University of California, Los Angeles, CA 90095, USA}\\
{\tt \small corbino@physics.ucla.edu}}

\date{\today}

\begin{titlingpage}
\maketitle

\begin{abstract}
We investigate the existence of solutions with 16 supersymmetries to Type IIB supergravity on a spacetime of the form $AdS_{2}\times S^{5}\times S^{1}$ warped over a two-dimensional Riemann surface $\Sigma$. The existence of the Lie superalgebra $SU(1,1|4) \subset PSU(2,2|4)$, whose maximal bosonic subalgebra is $SO(2,1)\oplus SO(6)\oplus SO(2)$, motivates the search for half-BPS solutions with this isometry that are asymptotic to $AdS_{5} \times S^{5}$. We reduce the BPS equations to the Ansatz for the bosonic fields and supersymmetry generators compatible with these symmetries, then show that the only non-trivial solution is the maximally supersymmetric solution $AdS_{5}\times S^{5}$. We argue that this implies that no solutions exist for fully back-reacted D7 probe or D7/D3 intersecting branes whose near-horizon limit is of the form $AdS_{2}\times S^{5}\times S^{1}\times \Sigma$.
\end{abstract}

\end{titlingpage}

\vfill\eject

\setcounter{tocdepth}{1} 

\baselineskip=15pt
\setcounter{equation}{0}
\setcounter{footnote}{0}

\section{Introduction}

In the study of half-BPS solutions to Type IIB supergravity, recent progress was made on spacetimes with the following factors warped over a two-dimensional Riemann surface $\Sigma$: $AdS_{6}\times S^{2}$ \cite{DHoker:2016ujz, DHoker:2016ysh, DHoker:2017mds, DHoker:2017zwj} and $AdS_{2} \times S^{6}$ \cite{Corbino:2017tfl, Corbino:2018fwb}. In the former case, globally regular and geodesically complete solutions were obtained which provide the near-horizon geometry of $(p,q)$ five-brane webs \cite{Aharony:1997ju,Aharony:1997bh}. Such solutions are holographic duals to five-dimensional superconformal field theories, with the $SO(2,5)\oplus SO(3)$ isometry extending to invariance under the exceptional Lie superalgebra $F(4)$. In the latter case, the isometry $SO(2,1)\oplus SO(7)$ extends to a different real form of the Lie superalgebra $F(4)$. While solutions were obtained which locally match those of $(p,q)$-strings \cite{Schwarz:1995dk} in the near-horizon limit, some questions remain regarding their geodesic completeness. Additional families of non-compact globally regular and geodesically complete solutions were obtained independent from any string junction interpretation.


These two cases provide the most recent example of half-BPS solutions to supergravity on pairs of spacetimes with internal factors related by ``double analytic continuation'', e.g. $AdS_{p} \times S^{q}$ and $AdS_{q} \times S^{p}$. While the Minkowski signature of the overall spacetime precludes the possibility of having more than one $AdS$ factor, one may in general have multiple internal spaces of Euclidean signature, provided the bosonic symmetries of the half-BPS configuration are appropriately realized. An earlier example where such pairs of half-BPS solutions to Type IIB supergravity have been constructed are the spacetimes $AdS_{4}\times S^{2}\times S^{2}\times \Sigma$ \cite{DHoker:2007zhm} and $AdS_{2}\times S^{2}\times S^{4}\times \Sigma$ \cite{DHoker:2007mci}, providing the holographic duals to interface solutions and Wilson loops (respectively). In both cases, the solutions are asymptotic to $AdS_{5}\times S^{5}$, and the respective isometries extend to invariance under Lie superalgebras that are both subalgebras with 16 fermionic generators of $PSU(2,2|4)$.


In \cite{DHoker:2008wvd}, a general correspondence was proposed between certain Lie superalgebras with 16 fermionic generators and half-BPS solutions to either Type IIB supergravity or M-theory. In the case of Type IIB, the semi-simple Lie superalgebras $\mathcal{H}$ are subalgebras of $PSU(2,2|4)$, and the corresponding half-BPS solutions are invariant under $\mathcal{H}$ and locally asymptotic to the maximally supersymmetric solution $AdS_{5}\times S^{5}$. It is shown that there exist a finite number of such subalgebras $\mathcal{H}$, and thus one obtains a classification of half-BPS solutions with the above asymptotics. Among these are the special classes of exact solutions previously found in \cite{DHoker:2007zhm} and \cite{DHoker:2007mci}, while those of \cite{DHoker:2016ujz, DHoker:2016ysh, DHoker:2017mds, DHoker:2017zwj} and \cite{Corbino:2017tfl, Corbino:2018fwb} are absent since neither $F(4)$ nor any of its real forms are subalgebras of $PSU(2,2|4)$.


Half-BPS solutions related to D7 branes in Type IIB supergravity are of particular interest. The near-horizon limits of D7 probe or D7/D3 intersecting branes seem to support the existence of corresponding half-BPS solutions. However, D7 branes also produce flavor multiplets which ultimately break exact conformal invariance, and by the arguments of \cite{Kirsch:2005uy} and \cite{Buchbinder:2007ar,Harvey:2008zz} (which follows earlier work on D7 branes in \cite{Aharony:1998xz,Grana:2001xn}) no fully back-reacted near-horizon limit solutions corresponding to D7 branes should exist. The classification of \cite{DHoker:2008wvd} reveals two cases, corresponding to the subalgebras $SU(1,1|4)\oplus SU(1,1)$ and $SU(2,2|2)\oplus SU(2)$, whose global symmetries and spacetime structure match those of the D7 probe or D7/D3 intersecting brane analysis. Each superalgebra contains a purely bosonic invariant subalgebra, respectively $SU(1,1)$ and $SU(2)$, which is not required by superconformal invariance. Additionally, these extra bosonic invariant subalgebras are incompatible with asymptotic $AdS_{5}\times S^{5}$ behavior (see Section 5.4 of \cite{DHoker:2008wvd} and references therein). Their removal yields the respective cases $SU(1,1|4)$ and $SU(2,2|2)$, for which the superalgebra correspondence suggests the existence of half-BPS solutions. However, these cases no longer possess the symmetries necessary for fully back-reacted near-horizon D7 brane solutions.


In this paper, we consider half-BPS solutions with $SO(2,1)\oplus SO(6)\oplus SO(2)$ symmetry, corresponding to the maximal bosonic symmetry of the superalgebra $SU(1,1|4)$ and realized on a spacetime of the form $AdS_{2}\times S^{5}\times S^{1}$ warped over a Riemann surface $\Sigma$. Half-BPS solutions invariant under $SU(2,2|2)$ were investigated in \cite{DHoker:2010gus}, where it was shown that on either $AdS_{5}\times S^{2} \times S^{1} \times \Sigma$ or $AdS_{5}\times S^{3} \times \Sigma$ the only non-trivial solution that exists is $AdS_{5}\times S^{5}$. Employing the same strategy here, we prove that the only half-BPS solution invariant under $SO(2,1)\oplus SO(6)\oplus SO(2)$ is once again $AdS_{5}\times S^{5}$. Thus, in the supergravity limit no fully back-reacted solutions of D7 branes can exist whose near-horizon limit match the symmetries and spacetime geometries of either case. Note that in contrast to the $SU(2,2|2)$ solutions, the bosonic invariant Lie subalgebra $SU(1,1)$ which is removed to obtain the $SU(1,1|4)$ case corresponds to a part of the isometry algebra for the original Anti-de Sitter space, and thus part of the conformal symmetry of the higher-dimensional dual CFT is broken. Additional arguments presented in Section 5.4 of \cite{DHoker:2008wvd} provide further evidence that the case of $SU(1,1|4)\oplus SU(1,1)$ cannot support half-BPS solutions with genuine asymptotic $AdS_{5}\times S^{5}$ behavior, and so we do not consider such solutions here.


\subsection{Organization}


The remainder of this paper is organized as follows. In Section \ref{sec:IIB}, we review Type IIB supergravity and introduce the general $SO(2,1)\oplus SO(6)\oplus SO(2)$-invariant Ansatz. In Section \ref{sec:reducingBPS}, we reduce the BPS equations to this Ansatz and examine the symmetries of these reduced equations. In Section \ref{sec:metricspinors}, we solve the reduced BPS equations for the metric factors of the $AdS_{2}\times S^{5}\times S^{1}$ spaces, and express them in terms of bilinears in the supersymmetry spinors. In Section \ref{sec:vanishinghermforms}, we obtain various sets of Hermitian relations implied by the reduced BPS equations. In Section \ref{sec:gensolnsBPS}, we solve these relations and show how they imply that the only solution with at least 16 supersymmetries for a spacetime of the form $AdS_{2}\times S^{5}\times S^{1}\times \Sigma$ is just the maximally supersymmetric solution $AdS_{5}\times S^{5}$. We conclude with a discussion in Section \ref{sec:conclusion}. In Appendix \ref{sec:Clifford}, a basis for the Clifford algebra adapted to the Ansatz is presented. Finally, in Appendix \ref{sec:BPS}, we discuss further details of the reduction of the BPS equations.






\section{ \texorpdfstring{$AdS_{2} \times S^{5} \times S^{1}\times \Sigma$}{AdS2xS5xS1xSigma} Ansatz in Type IIB supergravity}\label{sec:IIB} 

In this section, we review key aspects of Type IIB supergravity, then obtain the Ansatz for bosonic supergravity fields and susy generators with $SO(2,1)\oplus SO(6)\oplus SO(2)$ symmetry.

\subsection{Type IIB supergravity review}

The bosonic fields of Type IIB supergravity consist of the metric $g_{MN}$, the complex-valued axion-dilaton field $B$, a complex-valued  two-form potential $C_{(2)}$ and a real-valued four-form field $C_{(4)}$. The field strengths of the potentials $C_{(2)}$ and $C_{(4)}$ are given as follows,
\begin{align}
F_{(3)} & = d C_{(2)} & F_{(5)} & = dC_{(4)} + \frac{i}{16}(C_{(2)} \wedge \bar{F}_{(3)} - \bar{C}_{(2)} \wedge F_{(3)})
\end{align}
The field strength $F_{(5)}$ satisfies the well-known self-duality condition $F_{(5)} = *F_{(5)}$. Instead of the scalar field $B$ and the 3-form $F_{(3)}$, the fields that actually enter the BPS equations are composite fields, namely the one-forms $P,Q$ representing $B$, and the complex 3-form $G$ representing $F_{(3)}$, given in terms of the fields defined above by the following relations,
\begin{align}\label{eq:PQG}
P & = f_{B}^2 \, dB & f_{B}^2 & = ( 1 - |B|^2)^{-1} \nonumber \\
Q & = f_{B}^{2} \, \Im (B \, d\bar{B}) && \nonumber \\
G & = f_{B} (F_{(3)} - B\bar{F}_{(3)}) &&
\end{align}
Under the $SU(1,1)\sim SL(2,\mathds{R})$ global symmetry of Type IIB supergravity, the Einstein-frame metric $g_{MN}$ and the four-form $C_{(4)}$ are invariant, while $B$ and $C_{(2)}$ transform as,
\begin{align}\label{eq:globalsymm1} 
B & \to \frac{uB + v}{\bar{v}B + \bar{u}} & C_{(2)} & \to uC_{(2)} + v\bar{C}_{(2)}
\end{align}
where $SU(1,1)$ is parametrized by  $u, v \in \mathds{C}$ with $|u|^{2} - |v|^{2} = 1$. The field $B$ takes values in the coset $SU(1,1)/U(1)_{q}$ and $Q$ acts as a composite $U(1)_q$ gauge field. Given (\ref{eq:PQG}) and (\ref{eq:globalsymm1}), the $SU(1,1)$ symmetry induces the following transformations on the composite fields \cite{Schwarz:1983qr},
\begin{align}\label{eq:globalsymm2} 
P & \to e^{2i\theta}P & \theta & = \arg (v \bar{B} + u) \nonumber \\
Q & \to Q + d\theta && \nonumber \\
G & \to e^{i\theta}G &&
\end{align}
Equivalently, one may formulate Type IIB supergravity directly in terms of $g_{MN}$, $F_{(5)}$, $P,Q$ and $G$
provided these fields are subject to the Bianchi identities \cite{Schwarz:1983qr, Howe:1983sra},
\begin{align}\label{eq:Bianchi} 
0 & = dP - 2i Q \wedge P \\
0 & = dQ + i P\wedge \bar{P} \\
0 & = dG - i Q \wedge G + P \wedge \bar{G} \\
0 & = dF_{(5)} - \frac{i}{8} G \wedge \bar{G}
\end{align}
The fermion fields of Type IIB supergravity are the dilatino $\lambda$ and the gravitino $\psi_{M}$. The conditions that these fields and their variations $\delta \lambda$, $\delta \psi_{M}$ vanish yield the BPS equations, \footnote{Repeated indices are summed over, and complex conjugation is denoted by a \textit{bar} for functions and by a \textit{star} for spinors. We use the notation $\Gamma\cdot T \equiv \Gamma^{A_{1}\cdots A_{p}}T_{A_{1}\cdots A_{p}}$ for the contraction of an antisymmetric tensor field $T$ of rank $p$ with a $\Gamma$-matrix of the same rank. The matrices $\Gamma^{A}$ and $\mathcal{B}$ are defined in Appendix \ref{sec:Clifford}.} 
\begin{align}\label{eq:BPSeqns} 
0 & = i(\Gamma\cdot P)\mathcal{B}^{-1}\varepsilon^{*} - \frac{i}{24}(\Gamma\cdot G)\varepsilon \nonumber \\
0 & = (\nabla_{M} - { i \over 2} Q_M) \varepsilon + \frac{i}{480}(\Gamma\cdot F_{(5)})\Gamma_{M}\varepsilon - \frac{1}{96}(\Gamma_{M}(\Gamma\cdot G) + 2(\Gamma\cdot G)\Gamma_{M})\mathcal{B}^{-1}\varepsilon^{*}
\end{align}
where $\varepsilon$ is the supersymmetry generator transforming under the minus chirality Weyl spinor representation of $SO(1,9)$ and $\nabla _M$ is the covariant derivative acting on this representation.

\subsection{$SO(2,1)\oplus SO(6)\oplus SO(2)$-invariant Ansatz for supergravity fields}

We construct a general Ansatz for the bosonic fields of Type IIB supergravity consistent with the $SO(2,1)\oplus SO(6)\oplus SO(2)$ symmetry algebra. A natural realization is a spacetime geometry of the form $AdS_{2}\times S^{5}\times S^{1}$ warped over a two-dimensional Riemann surface $\Sigma$. The $SO(2,1)\oplus SO(6)\oplus SO(2)$-invariant Ansatz for the metric is then of the following form,
\begin{align}
ds^{2} = f_{2}^{2}\,d\hat{s}_{AdS_{2}}^{2} + f_{5}^{2}\,d\hat{s}_{S^{5}}^{2} + f_{1}^{2}\,d\hat{s}_{S^{1}}^{2} + ds_{\Sigma}^{2}
\end{align}
where the radii $f_{2}, f_{5}, f_{1}$ and $ds_{\Sigma}^{2}$ are functions of $\Sigma$. We define an orthonormal frame,
\begin{align}\label{eq:frames} 
e^{m} & = f_{2} \, \hat{e}^{m} & m & = 0,1 \nonumber \\
e^{i} & = f_{5} \, \hat{e}^{i} & i & =  2, 3, 4, 5, 6 \nonumber \\
e^{a} & = \rho \,\, \hat{e}^{a} & a & = 7, 8 \nonumber \\
e^{9} & = f_{1} \, \hat{e}^{9} &&
\end{align}
where $\hat{e}^{m}$, $\hat{e}^{i}$, and $\hat{e}^{9}$ respectively refer to orthonormal frames for the spaces $AdS_{2}$, $S^{5}$, and $S^{1}$ with unit radius. Here, ${e}^{a}$ is an orthonormal frame on $\Sigma$ only, so that we have,
\begin{align}
d\hat s^{2}_{AdS_{2}} &= \eta^{(2)}_{mn}\,\hat{e}^{m} \otimes \hat{e}^{n} & d\hat s^{2}_{S^{5}} &= \delta_{ij}\,\hat{e}^{i} \otimes \hat{e}^{j} \nonumber \\
ds^{2}_{\Sigma} &= \delta_{ab}\, e^{a} \otimes  e^{b} & d\hat s^{2}_{S^{1}} &= \hat{e}^{9} \otimes \hat{e}^{9}
\end{align}
The axion-dilaton field $B$ is a function of $\Sigma$ only, so the 1-forms $P$ and $Q$ can be written as,
\begin{align}
P & = p_{a}e^{a} & Q & = q_{a}e^{a}
\end{align}
where the components $p_{a}$, $q_{a}$ are complex and depend on $\Sigma$ only. Finally, the complex 3-form $G$ and self-dual 5-form field strength $F_{(5)} = * F_{(5)}$ are given as follows,
\begin{align}
G & = i g_{\bar{a}}e^{01 \bar{a}} + h e^{789} & F_{(5)} & = f \left( e^{01789} + e^{23456} \right)
\end{align}
where the indices $\bar{a}$ run over the values $7,8,9$. 
The coefficients are constrained by $SO(2,1)\oplus SO(6)\oplus SO(2)$ invariance, so that both the real-valued functions $f$, $q_{a}$ and complex-valued functions $p_{a}$, $h$, $g_{a}$, $g_{9}$ depend only on $\Sigma$. This completes the Ansatz for the bosonic fields.

\subsection{$SO(2,1)\oplus SO(6)\oplus SO(2)$-invariant Ansatz for susy generators}

We decompose the supersymmetry generator $\varepsilon$ onto the Killing spinors of the various components of $AdS_{2}\times S^{5}\times S^{1}$. The Killing spinor equations on $AdS_{2}$ and on $S^{5}$ were derived in the appendices of \cite{DHoker:2007mci} and \cite{DHoker:2008lup}, and are given (respectively) by,
\begin{align}
\label{eq:KSeqns} 
\Big ( \hat{\nabla}_{m} - \frac{1}{2}\eta_{1} \gamma_{m} \otimes I_{4} \Big ) \chi^{\eta_{1},\eta_{2}} & = 0 & \Big ( \hat{\nabla}_{i} - \frac{i}{2}\eta_{2} I_{4} \otimes \gamma_{i} \Big ) \chi^{\eta_{1},\eta_{2}} & = 0
\end{align}
Here, $\hat{\nabla}_{m}$ and $\hat{\nabla}_{i}$ are the covariant spinor derivatives on the respective spaces, and integrability requires that $\eta_{1}^{2} = \eta_{2}^{2} = 1$. The action of the chirality matrices is given by,
\begin{align}
\left( \gamma_{(1)} \otimes I_{8} \right) \chi^{\eta_{1},\eta_{2}} & = \chi^{-\eta_{1},\eta_{2}} & \left( I_{2} \otimes \gamma_{(2)} \right) \chi^{\eta_{1},\eta_{2}} & = \chi^{\eta_{1},+\eta_{2}}
\end{align}
while under charge conjugation we have,
\begin{align}
\chi^{\eta_{1},\eta_{2}} \to \left( \chi^{c} \right)^{\eta_{1},\eta_{2}} = \left( B_{(1)} \otimes B_{(2)} \right)^{-1}\left( \chi^{\eta_{1},\eta_{2}} \right)^{*} \propto \chi^{-\eta_{1},-\eta_{2}}
\end{align}
The components are found by first choosing $\left( \chi^{c} \right)^{++} \equiv \chi^{--}$, then using the chirality matrix $\gamma_{(1)}$ and charge conjugation matrices $B_{(1)}, B_{(2)}$ to obtain the following relations for all $\eta_{1},\eta_{2}$:
\begin{align}
\left( \chi^{c} \right)^{\eta_{1},\eta_{2}}  = \eta_{2}\chi^{-\eta_{1},-\eta_{2}}
\end{align}
Killing spinors $\chi^{\eta_{3}}$ on $S^{1}$ are single functions for each value of $\eta_{3}$ which solve the equation,
\begin{align}
\left( \hat{\nabla}_{9} - \frac{i}{2}\eta_{3} \right)\chi^{\eta_{3}} = 0
\end{align}
As explained in \cite{DHoker:2010gus}, we may set $(\chi^{\eta_{3}})^{*} = \chi^{-\eta_{3}}$, with the values $\eta_{3} = \pm 1$ corresponding to a double-valued representation for the spinors. The most general $32$-component complex spinor $\varepsilon$ that can be decomposed onto the Killing spinors of $AdS_{2}\times S^{5}\times S^{1}$, and which is consistent with the 10-dimensional chirality condition $\Gamma^{11}\varepsilon = -\varepsilon$, is of the following form,
\begin{align}\label{eq:complexspinor} 
\varepsilon = \sum_{\eta_{1},\eta_{2},\eta_{3}}\chi^{\eta_{1},\eta_{2}}\chi^{\eta_{3}}\otimes\zeta_{\eta_{1},\eta_{2},\eta_{3}}\otimes\phi
\end{align}
where we have defined the constant spinor,
\begin{align}
\phi \equiv e^{-i\pi/4}
\begin{pmatrix} 
1 \\
0 \\
\end{pmatrix}
+ e^{i\pi/4}
\begin{pmatrix} 
0 \\
1 \\
\end{pmatrix}
\end{align}
Finally, the charge conjugate spinor is given by,
\begin{align}
\mathcal{B}^{-1}\varepsilon^{*} & = \sum_{\eta_{1},\eta_{2},\eta_{3}}\chi^{\eta_{1},\eta_{2}}\chi^{\eta_{3}}\otimes\star\zeta_{\eta_{1},\eta_{2},\eta_{3}}\otimes\phi & \star\zeta_{\eta_{1},\eta_{2},\eta_{3}} & = -i\eta_{2}\sigma^{2}\zeta^{*}_{-\eta_{1},-\eta_{2},-\eta_{3}}
\end{align}
This completes the construction of the $SO(2,1) \oplus SO(6) \oplus SO(2)$-invariant Ansatz.

\section{Reducing the BPS equations}\label{sec:reducingBPS} 


The residual supersymmetries, if any, of a configuration of purely bosonic Type IIB supergravity fields are governed by the BPS equations of (\ref{eq:BPSeqns}). In \cite{DHoker:2016ujz, DHoker:2016ysh, DHoker:2017mds, DHoker:2017zwj} and \cite{Corbino:2017tfl, Corbino:2018fwb}, the Ansatz for the supergravity fields and supersymmetry spinor which satisfies the BPS equations also solves the Bianchi identities and field equations, and thus automatically provides a half-BPS solution to Type IIB supergravity. In the present case however, we will show that no non-trivial half-BPS solutions exist. In this section, we reduce the BPS equations to the $AdS_{2}\times S^{5}\times S^{1}\times \Sigma$ Ansatz, expose its residual symmetries, and utilize the complex structure on the Riemann surface to separate the chirality components of the 2-dimensional spinors $\zeta$.

\subsection{The reduced BPS equations}

As in \cite{DHoker:2016ujz,Corbino:2017tfl} we use the $\tau$ matrix notation introduced originally in \cite{Gomis:2006cu} to compactly express the action of the various $\gamma$ matrices on $\zeta$. Defining $\tau^{(ijk)} = \tau^{i}\otimes\tau^{j}\otimes\tau^{k}$ with $i,j,k = 0, 1, 2, 3$, $\tau^{0}$ the identity matrix and $\tau^{i}$ for $i = 1, 2, 3$ the standard Pauli matrices, we can write,
\begin{align}
( \tau^{(ijk)}\zeta )_{\eta_{1}^{},\eta_{2}^{},\eta_{3}^{}} \equiv \sum_{\eta_{1}',\eta_{2}',\eta_{3}'}(\tau^{i})_{\eta_{1}^{}\eta_{1}'}(\tau^{j})_{\eta_{2}^{}\eta_{2}'}(\tau^{k})_{\eta_{3}^{}\eta_{3}'}\zeta_{\eta_{1}',\eta_{2}',\eta_{3}'}
\end{align}
The reduction of the BPS equations (\ref{eq:BPSeqns}) using the decomposition of $\varepsilon$ (\ref{eq:complexspinor}) onto the Killing spinors (\ref{eq:KSeqns}) is discussed in Appendix \ref{sec:BPS}. The reduced dilatino BPS equation is given by,
\begin{align}\label{eq:dilatino} 
& (d) & 0 & = 4p_{a}\gamma^{a}\sigma^{2}\zeta^{*} + i g_{\bar{a}}\tau^{(021)}\gamma^{\bar{a}}\zeta - i h\tau^{(121)}\zeta
\end{align}
while the various components of the reduced gravitino BPS equations are as follows,
\begin{align}\label{eq:gravitino} 
& (m) & 0 & = \frac{1}{2f_{2}}\tau^{(300)}\zeta + \frac{D_{a}f_{2}}{2f_{2}}\tau^{(100)}\gamma^{a}\zeta + \frac{1}{2}f\zeta \nonumber \\
&&& \quad + \frac{1}{16}\left( 3ig_{\bar{a}}\tau^{(121)}\gamma^{\bar{a}}\sigma^{2}\zeta^{*} + ih\tau^{(021)}\sigma^{2}\zeta^{*} \right) \nonumber \\
& (i) & 0 & = \frac{1}{2f_{5}}\tau^{(030)}\zeta + \frac{D_{a}f_{5}}{2f_{5}}\tau^{(100)}\gamma^{a}\zeta - \frac{1}{2}f\zeta \nonumber \\
&&& \quad + \frac{1}{16}\left( -ig_{\bar{a}}\tau^{(121)}\gamma^{\bar{a}}\sigma^{2}\zeta^{*} + ih\tau^{(021)}\sigma^{2}\zeta^{*} \right) \nonumber \\
& (a) & 0 & = \left( D_{a} + \frac{i}{2}\hat{\omega}_{a}\sigma^{3} \right)\zeta - \frac{i}{2}q_{a}\zeta + \frac{1}{2}f\tau^{(100)}\gamma_{a}\zeta \nonumber \\
&&& \quad + \frac{1}{16}\left( 3ig_{a}\tau^{(021)}\sigma^{2}\zeta^{*} - ig_{\bar{b}}\tau^{(021)}\gamma_{a}{}^{\bar{b}}\sigma^{2}\zeta^{*} - 3ih\tau^{(121)}\gamma_{a}\sigma^{2}\zeta^{*} \right) \nonumber \\
& (9) & 0 & = \frac{i}{2f_{1}}\tau^{(103)}\sigma^{3}\zeta + \frac{D_{a}f_{1}}{2f_{1}}\tau^{(100)}\gamma^{a}\zeta + \frac{1}{2}f\zeta \nonumber \\
&&& \quad + \frac{1}{16}\left( 3ig_{9}\tau^{(121)}\sigma^{3}\sigma^{2}\zeta^{*} - ig_{a}\tau^{(121)}\gamma^{a}\sigma^{2}\zeta^{*} - 3ih\tau^{(021)}\sigma^{2}\zeta^{*} \right)
\end{align}
The derivative $D_{a}$ is defined with respect to the frame $e^{a}$ of $\Sigma$, so that the total differential $d_\Sigma$ takes the form $d_\Sigma  = e^{a}D_{a}$, while the $U(1)$-connection with respect to frame indices is $\hat{\omega}_{a}$.

\subsection{Symmetries of the reduced BPS equations}

The global $SU(1,1)$ symmetry of Type IIB, whose action on the bosonic fields was given in (\ref{eq:globalsymm1}) and (\ref{eq:globalsymm2}), survives the reduction to the $SO(2,1)\oplus SO(6) \oplus SO(2)$-invariant Ansatz. 
Upon reduction to the Ansatz, the $U(1)_{q}$ gauge transformations of (\ref{eq:globalsymm2}) are now given by,
\begin{align}
\zeta & \to e^{i\theta/2}\zeta & g_{a} & \to e^{i\theta}g_{a} \nonumber \\
q_{a} & \to q_{a} + D_{a}\theta & g_{9} & \to e^{i\theta}g_{9} \nonumber \\
p_{a} & \to e^{2i\theta}p_{a} & h & \to e^{i\theta}h
\end{align}
In addition to the continuous symmetries, there are linear discrete symmetries which leave the reduced supergravity fields unchanged and act on the supersymmetry generator as follows,
\begin{align}\label{eq:discretesymm} 
\zeta \to \zeta' & = S\zeta & S \in \mathcal{S}_{0} & \equiv \left\{ I, \tau^{(033)}, i\tau^{(030)}, i\tau^{(003)} \right\}
\end{align}
Finally, composing complex conjugation with an arbitrary $U(1)_{q}$ transformation, we have,
\begin{align}\label{eq:CCsymm} 
\zeta & \to \mathcal{K}\zeta = e^{i\theta}\tau^{(033)}\sigma^{1}\zeta^{*} & g_{a} & \to \mathcal{K}g_{a} = e^{2i\theta}g_{a}^{*} \nonumber \\
p_{a} & \to \mathcal{K}p_{a} = e^{4i\theta}p_{a}^{*} & g_{9} & \to \mathcal{K}g_{9} = -e^{2i\theta}g_{9}^{*} \nonumber \\
q_{a} & \to \mathcal{K}q_{a} = -q_{a} + 2D_{a}\theta & h & \to \mathcal{K}h = e^{2i\theta}h^{*}
\end{align}
while the pure discrete complex conjugation corresponds to the special case where $\theta = 0$.

\subsection{Further reduction and chiral form of the BPS equations}

We now derive the restrictions to one of the linear discrete symmetries which are implied by the reduced BPS equations, following the same procedure that was used for \cite{DHoker:2010gus}. 
From (\ref{eq:discretesymm}), we see that only $\tau^{(033)}\in\mathcal{S}_{0}$ commutes with the BPS differential operator and admits real eigenvalues. Therefore, we may diagonalize this symmetry simultaneously with the BPS operator and analyze separately the restriction of the BPS equations to the two eigenspaces,
\begin{align}
\zeta \to \tau^{(033)}\zeta & = \nu\zeta & \nu & = \pm 1
\end{align}
The non-zero components of $\zeta$ are then redefined in terms of a new $\zeta$-spinor with two indices,
\begin{align}\label{eq:twoindexzeta} 
\nu & = \pm 1 : & \zeta_{\eta_{1},\eta_{2}} & \equiv \zeta_{\eta_{1},\eta_{2},\nu\eta_{2}}
\end{align}
The remaining elements $i\tau^{(030)}, i\tau^{(003)}\in \mathcal{S}_{0}$ (\ref{eq:discretesymm}) map between identical $\nu$, and along with the complex conjugations symmetries (\ref{eq:CCsymm}) reduce under (\ref{eq:twoindexzeta}) to the following transformations,
\begin{align}\label{eq:reducedCCsymm} 
i\tau^{(030)}\zeta & \to i\tau^{(03)}\zeta & i\tau^{(003)}\zeta & \to i\nu\tau^{(03)}\zeta &\mathcal{K}\zeta \to \nu e^{i\theta}\sigma^{1}\zeta^{*}
\end{align}
%
%
We then decompose the spinors $\zeta_{\eta_{1},\eta_{2}}$ in terms of complex frame basis $e^{a} = (e^{z},e^{\bar{z}})$ on $\Sigma$, with a metric $\delta_{z\bar{z}} = \delta_{\bar{z}z} = 2$ and Clifford algebra generators $\gamma^{a} = (\gamma^{z},\gamma^{\bar{z}})$ defined as follows,
\begin{align}
e^{z} & = \frac{1}{2}\left( e^{7} + i e^{8} \right) & e^{\bar{z}} & = \frac{1}{2}\left( e^{7} - i e^{8} \right) &
\gamma^{z} & =
\begin{pmatrix}
0 & 1\\
0 & 0
\end{pmatrix}
& \gamma^{\bar{z}} & =
\begin{pmatrix}
0 & 0\\
1 & 0
\end{pmatrix}
\end{align}
Similar relations hold for $p_{a}, q_{a}, g_{a}$, e.g. $p_{z} = p_{7} - ip_{8}$ and $p_{\bar{z}} = p_{7} + ip_{8}$. In this same 2-dimensional spinor basis, we decompose the two-index spinor $\zeta$ into the chirality components,
\begin{align}
\zeta_{\eta_{1},\eta_{2}} & = 
\begin{pmatrix}
\tau^{(02)}\xi_{\eta_{1},\eta_{2}}^{*} \\
\psi_{\eta_{1},\eta_{2}}
\end{pmatrix}
\end{align}
where $\xi_{\eta_{1},\eta_{2}}^{*}$, $\psi_{\eta_{1},\eta_{2}}$ are 1-component spinors. In this basis, the reduced dilatino equation is,
\begin{align}\label{eq:reducedD} 
& (d_{1}) & 0 & = 4ip_{z}\xi - ig_{z}\psi - ig_{9}\tau^{(02)}\xi^{*} + ih\tau^{(12)}\xi^{*} \nonumber \\
& (d_{2}) & 0 & = 4ip_{\bar{z}}^{*}\psi - ig_{\bar{z}}^{*}\xi - ig_{9}^{*}\tau^{(02)}\psi^{*} - ih\tau^{(12)}\psi^{*}
\end{align}
The components of the reduced gravitino equation along $AdS_{2}$, $S^{5}$, $S^{1}$ are given by,
\begin{align}\label{eq:reducedGads} 
& (m_{1}) & 0 & = \frac{-i}{2f_{2}}\tau^{(22)}\xi^{*}+\frac{D_{z}f_{2}}{2f_{2}}\psi + \frac{1}{2}f\tau^{(12)}\xi^{*} + \frac{1}{16} \left( 3g_{z}\xi + 3g_{9}\tau^{(02)}\psi^{*} + h\tau^{(12)}\psi^{*} \right) \nonumber \\
& (m_{2}) & 0 & = \frac{i}{2f_{2}}\tau^{(22)}\psi^{*}+\frac{D_{z}f_{2}}{2f_{2}}\xi - \frac{1}{2}f\tau^{(12)}\psi^{*} + \frac{1}{16} \left( 3g_{\bar{z}}^{*}\psi + 3g_{9}^{*}\tau^{(02)}\xi^{*} - h^{*}\tau^{(12)}\xi^{*} \right) \nonumber \\
& (i_{1}) & 0 & = \frac{-i}{2f_{5}}\tau^{(11)}\xi^{*}+\frac{D_{z}f_{5}}{2f_{5}}\psi - \frac{1}{2}f\tau^{(12)}\xi^{*} + \frac{1}{16} \left( -g_{z}\xi - g_{9}\tau^{(02)}\psi^{*} + h\tau^{(12)}\psi^{*} \right) \nonumber \\
& (i_{2}) & 0 & = \frac{-i}{2f_{5}}\tau^{(11)}\psi^{*}+\frac{D_{z}f_{5}}{2f_{5}}\xi + \frac{1}{2}f\tau^{(12)}\psi^{*} + \frac{1}{16} \left( -g_{\bar{z}}^{*}\psi - g_{9}^{*}\tau^{(02)}\xi^{*} - h^{*}\tau^{(12)}\xi^{*} \right) \nonumber \\
& (9_{1}) & 0 & = \frac{\nu}{2f_{1}}\tau^{(01)}\xi^{*}+\frac{D_{z}f_{1}}{2f_{1}}\psi + \frac{1}{2}f\tau^{(12)}\xi^{*} + \frac{1}{16} \left( -g_{z}\xi + 3g_{9}\tau^{(02)}\psi^{*} - 3h\tau^{(12)}\psi^{*} \right) \nonumber \\
& (9_{2}) & 0 & = \frac{\nu}{2f_{1}}\tau^{(01)}\psi^{*}+\frac{D_{z}f_{1}}{2f_{1}}\xi - \frac{1}{2}f\tau^{(12)}\psi^{*} + \frac{1}{16} \left( -g_{\bar{z}}^{*}\psi + 3g_{9}^{*}\tau^{(02)}\xi^{*} + 3h^{*}\tau^{(12)}\xi^{*} \right)
\end{align}
together with the components along $\Sigma$,
\begin{align}\label{eq:reducedGsigma} 
& (+_{1}) & 0 & = \left( D_{\bar{z}} - \frac{i}{2}\hat{\omega}_{\bar{z}} + \frac{i}{2}q_{\bar{z}} \right)\xi + \frac{1}{4}g_{z}^{*}\psi \nonumber \\
& (+_{2}) & 0 & = \left( D_{z} - \frac{i}{2}\hat{\omega}_{z} - \frac{i}{2}q_{z} \right)\psi + f\tau^{(12)}\xi^{*} + \frac{1}{8}\left( g_{z}\xi - g_{9}\tau^{(02)}\psi^{*} - 3h\tau^{(12)}\psi^{*} \right) \nonumber \\
& (-_{1}) & 0 & = \left( D_{z} - \frac{i}{2}\hat{\omega}_{z} + \frac{i}{2}q_{z} \right)\xi - f\tau^{(12)}\psi^{*} + \frac{1}{8}\left( g_{\bar{z}}^{*}\psi - g_{9}^{*}\tau^{(02)}\xi^{*} + 3h^{*}\tau^{(12)}\xi^{*} \right) \nonumber \\
& (-_{2}) & 0 & = \left( D_{\bar{z}} - \frac{i}{2}\hat{\omega}_{\bar{z}} - \frac{i}{2}q_{\bar{z}} \right)\psi + \frac{1}{4}g_{\bar{z}}\xi
\end{align}
where $\hat{\omega}_{z} = i(\partial_{w}\rho)/\rho^{2}$.
The action of the complex conjugation symmetry (\ref{eq:reducedCCsymm}) is given by,
\begin{align}
\xi \to \xi' & = -\nu e^{-i\theta}\tau^{(02)}\psi & \psi \to \psi' & = -\nu e^{+i\theta}\tau^{(02)}\xi
\end{align}
with the transformations on the bosonic fields (\ref{eq:CCsymm}) translated as follows in the chiral basis,
\begin{align}
p_{z} \to p_{z}' & = e^{4i\theta}(p_{\bar{z}})^{*} & g_{z} \to g_{z}' & = e^{2i\theta}(g_{\bar{z}})^{*} & h \to h' & = e^{2i\theta}h^{*} \nonumber \\
q_{z} \to q_{z}' & = -q_{z} + 2D_{z}\theta & g_{9} \to g_{9}' & = -e^{2i\theta}g_{9}^{*} & &
\end{align}
Finally, we note that shifting the metric factor $f_{1}\to \nu f_{1}$ removes all explicit dependence on $\nu$ from the reduced BPS equations, which is irrelevant since the supergravity fields only ever depend on the square $f_{1}^{2}$. Thus, for every solution to the reduced BPS equations with $\nu = +1$, there exists another solution with $\nu = -1$ so that a systematic doubling of the total number of spinor solutions is produced. Together with the counting of components for the basis of Killing spinors in (\ref{eq:complexspinor}), this implies that any solution with $\nu = +1$ produces 16 linearly independent solutions to the BPS equations, thereby generating a half-BPS solution.

\section{Metric factors in terms of spinor bilinears}\label{sec:metricspinors} 


In this section, we use the gravitino BPS equations to solve for the metric factors $f_{1}, f_{2}, f_{5}$. We find that their solutions may be related to bilinears of the spinors $\psi, \xi$. The reality properties of the metric factors impose the conditions that the spinor bilinears be real and invariant under $U(1)_{q}$ transformations. The only combinations that satisfy these requirements are those of the form $\psi^{\dagger}\tau^{(\alpha\beta)}\psi$, $\xi^{\dagger}\tau^{(\alpha\beta)}\xi$. We seek relations that hold for \textit{generic} values of the supergravity fields $f_{1},f_{2},f_{5},f,g_{z},g_{\bar{z}},h$ and $g_{9}$. Following the same procedure that was used for \cite{DHoker:2010gus}, we will use combinations of the differential equations $(\pm)$ in (\ref{eq:reducedGsigma}),
\begin{align}
D_{z}\left( \psi^{\dagger}\tau^{(\alpha\beta)}\psi \right) = & -f\psi^{\dagger}\tau^{(\alpha\beta)}\tau^{(12)}\xi^{*} - \frac{1}{4}g_{\bar{z}}^{*}\xi^{\dagger}\tau^{(\alpha\beta)}\psi \nonumber \\
& + \frac{1}{8}\left( -g_{z}\psi^{\dagger}\tau^{(\alpha\beta)}\xi + g_{9}\psi^{\dagger}\tau^{(\alpha\beta)}\tau^{(02)}\psi^{*} + 3h\psi^{\dagger}\tau^{(\alpha\beta)}\tau^{(12)}\psi^{*} \right) \nonumber \\
D_{z}\left( \xi^{\dagger}\tau^{(\alpha\beta)}\xi \right) = & +f\xi^{\dagger}\tau^{(\alpha\beta)}\tau^{(12)}\psi^{*} - \frac{1}{4}g_{z}\psi^{\dagger}\tau^{(\alpha\beta)}\xi \nonumber \\
& + \frac{1}{8}\left( -g_{\bar{z}}^{*}\xi^{\dagger}\tau^{(\alpha\beta)}\psi + g_{9}^{*}\xi^{\dagger}\tau^{(\alpha\beta)}\tau^{(02)}\xi^{*} - 3h^{*}\xi^{\dagger}\tau^{(\alpha\beta)}\tau^{(12)}\xi^{*} \right)
\end{align}
and of the algebraic gravitino BPS equations (\ref{eq:reducedGads}) to find relations of the following type,
\begin{align}
D_{z}\left( r_{1}\psi^{\dagger}\tau^{(\alpha\beta)}\psi + r_{2}\xi^{\dagger}\tau^{(\alpha\beta)}\xi \right) + \frac{D_{z}f_{i}}{f_{i}}\left( r_{3}\psi^{\dagger}\tau^{(\alpha\beta)}\psi + r_{4}\xi^{\dagger}\tau^{(\alpha\beta)}\xi \right) = 0
\end{align}
where $i= 1,2,5$, and the coefficients $r_{1},r_{2},r_{3},r_{4}$ may depend on $i$ and $\alpha,\beta$, but not on $\Sigma$.

%
%

\subsection{$AdS_{2}$ metric factor}
Left-multiplying equation $(m_{1})$ by $\psi^{\dagger}$ and the complex conjugate of equation $(m_{2})$ by $\xi^{\dagger}$, then term by term cancellation imposes the following requirements for generic fields,
\begin{align}
(f_{2}) & & 0 & = r_{3}\tau^{(\alpha\beta)}\tau^{(22)} - r_{4}\tau^{(22)}\left( \tau^{(\alpha\beta)} \right)^{t} \nonumber \\
(f) & & 0 & = (r_{1} + r_{3})\tau^{(\alpha\beta)}\tau^{(12)} + (r_{2} + r_{4})\tau^{(12)}\left( \tau^{(\alpha\beta)} \right)^{t} \nonumber \\
(g_{\bar{z}}^{*}) & & 0 & = 2r_{1} + r_{2} + 3r_{4} \nonumber \\
(g_{z}) & & 0 & = 2r_{2} + r_{1} + 3r_{3} \nonumber \\
(g_{9}) & & 0 & = (r_{1} - 3r_{3})g_{9}\psi^{\dagger}\tau^{(\alpha\beta)}\tau^{(02)}\psi^{*} = (r_{2} - 3r_{4})g_{9}^{*}\xi^{\dagger}\tau^{(\alpha\beta)}\tau^{(02)}\xi^{*} \nonumber \\
(h) & & 0 & = (3r_{1} - r_{3})h\psi^{\dagger}\tau^{(\alpha\beta)}\tau^{(12)}\psi^{*} = -(3r_{2} - r_{4})h^{*}\xi^{\dagger}\tau^{(\alpha\beta)}\tau^{(02)}\xi^{*}
\end{align}
If $r_{3} = 0$, then $(f_{2})$ implies $r_{4} = 0$, and $(g_{z})$ and $(g_{\bar{z}}^{*})$ imply $r_{1} = r_{2} = 0$. Therefore, for non-trivial solutions we have $r_{3}\neq 0$, and without loss of generality we set $r_{3} = 1$. Then $(f_{2})$ implies that $|r_{4}| = 1$, so that $(g_{z})$ and $(g_{\bar{z}}^{*})$ reduce to $r_{1} = -2r_{2} - 3$ and $r_{4} = r_{2} + 2$, with the condition that $|r_{2} + 1| = 1$. For $r_{4} = \pm 1$, $(f_{2})$ and $(f)$ yield two sets of solutions,
\begin{align}
(r_{1},r_{2},r_{3},r_{4}) & = (-1,-1,+1,+1) & \tau^{(\alpha\beta)} & \in \left\{ \tau^{(00)}, \tau^{(11)}, \tau^{(12)}, \tau^{(13)}, \tau^{(21)}, \right. \nonumber \\
&&& \qquad \left. \tau^{(22)}, \tau^{(23)}, \tau^{(31)}, \tau^{(32)}, \tau^{(33)} \right\} \nonumber \\
(r_{1},r_{2},r_{3},r_{4}) & = (+3,-3,+1,-1) & \tau^{(\alpha\beta)} & \in \left\{ \tau^{(10)}, \tau^{(20)} \right\}
\end{align}

\subsection{$S^{5}$ metric factor}

Left-multiplying equation $(i_{1})$ by $\psi^{\dagger}$ and the complex conjugate of equation $(i_{2})$ by $\xi^{\dagger}$, then term by term cancellation imposes the following requirements for generic fields,
\begin{align}
(f_{5}) & & 0 & = r_{3}\tau^{(\alpha\beta)}\tau^{(11)} + r_{4}\tau^{(11)}\left( \tau^{(\alpha\beta)} \right)^{t} \nonumber \\
(f) & & 0 & = (r_{3} - r_{1})\tau^{(\alpha\beta)}\tau^{(12)} + (r_{4} - r_{2})\tau^{(12)}\left( \tau^{(\alpha\beta)} \right)^{t} \nonumber \\
(g_{\bar{z}}^{*}) & & 0 & = 2r_{1} + r_{2} - r_{4} \nonumber \\
(g_{z}) & & 0 & = 2r_{2} + r_{1} - r_{3} \nonumber \\
(g_{9}) & & 0 & = (r_{1} + r_{3})g_{9}\psi^{\dagger}\tau^{(\alpha\beta)}\tau^{(02)}\psi^{*} = (r_{2} + r_{4})g_{9}^{*}\xi^{\dagger}\tau^{(\alpha\beta)}\tau^{(02)}\xi^{*} \nonumber \\
(h) & & 0 & = (3r_{1}-r_{3})h\psi^{\dagger}\tau^{(\alpha\beta)}\tau^{(12)}\psi^{*} = -(3r_{2} - r_{4})h^{*}\xi^{\dagger}\tau^{(\alpha\beta)}\tau^{(02)}\xi^{*}
\end{align}
If $r_{3} = 0$, then $(f_{5})$ implies $r_{4} = 0$, and $(g_{z})$ and $(g_{\bar{z}}^{*})$ imply $r_{1} = r_{2} = 0$. Therefore, for non-trivial solutions we have $r_{3}\neq 0$, and without loss of generality we set $r_{3} = 1$. Then $(f_{5})$ implies that $|r_{4}| = 1$, so that $(g_{z})$ and $(g_{\bar{z}}^{*})$ reduce to $r_{1} = -2r_{2} + 1$ and $r_{4} = -3r_{2} + 2$, with the condition that $|-3r_{2} + 2| = 1$. For $r_{4} = \pm 1$, $(f_{5})$ and $(f)$ yield two sets of solutions,
\begin{align}
(r_{1},r_{2},r_{3},r_{4}) & = (-1,+1,+1,-1) & \tau^{(\alpha\beta)} & \in \left\{ \tau^{(00)}, \tau^{(10)}, \tau^{(20)}, \tau^{(33)} \right\} \nonumber \\
(r_{1},r_{2},r_{3},r_{4}) & = (1/3,1/3,+1,+1) & \tau^{(\alpha\beta)} & \in \left\{ \tau^{(03)}, \tau^{(13)}, \tau^{(23)}, \tau^{(30)} \right\}
\end{align}

\subsection{$S^{1}$ metric factor}

Left-multiplying equation $(9_{1})$ by $\psi^{\dagger}$ and the complex conjugate of equation $(9_{2})$ by $\xi^{\dagger}$, then term by term cancellation imposes the following requirements for generic fields,
\begin{align}
(f_{1}) & & 0 & = r_{3}\tau^{(\alpha\beta)}\tau^{(01)} + r_{4}\tau^{(01)}\left( \tau^{(\alpha\beta)} \right)^{t} \nonumber \\
(f) & & 0 & = (r_{1}+r_{3})\tau^{(\alpha\beta)}\tau^{(12)} + (r_{2}+r_{4})\tau^{(12)}\left( \tau^{(\alpha\beta)} \right)^{t} \nonumber \\
(g_{\bar{z}}^{*}) & & 0 & = 2r_{1} + r_{2} - r_{4} \nonumber \\
(g_{z}) & & 0 & = 2r_{2} + r_{1} - r_{3} \nonumber \\
(g_{9}) & & 0 & = (r_{1}-3r_{3})g_{9}\psi^{\dagger}\tau^{(\alpha\beta)}\tau^{(02)}\psi^{*} = (r_{2} - 3r_{4})g_{9}^{*}\xi^{\dagger}\tau^{(\alpha\beta)}\tau^{(02)}\xi^{*} \nonumber \\
(h) & & 0 & = (r_{1}+r_{3})h\psi^{\dagger}\tau^{(\alpha\beta)}\tau^{(12)}\psi^{*} = -(r_{2} + r_{4})h^{*}\xi^{\dagger}\tau^{(\alpha\beta)}\tau^{(02)}\xi^{*}
\end{align}
If $r_{3} = 0$, then $(f_{1})$ implies $r_{4} = 0$, and $(g_{z})$ and $(g_{\bar{z}}^{*})$ imply $r_{1} = r_{2} = 0$. Therefore, for non-trivial solutions we have $r_{3}\neq 0$, and without loss of generality we set $r_{3} = 1$. Then $(f_{1})$ implies that $|r_{4}| = 1$, so that $(g_{z})$ and $(g_{\bar{z}}^{*})$ reduce to $r_{1} = -2r_{2} + 1$ and $r_{4} = -3r_{2} + 2$, with the condition that $|-3r_{2} + 2| = 1$. For $r_{4} = \pm 1$, $(f_{1})$ and $(f)$ yield two sets of solutions,
\begin{align}
(r_{1},r_{2},r_{3},r_{4}) & = (-1,+1,+1,-1) & \tau^{(\alpha\beta)} & \in \left\{ \tau^{(00)}, \tau^{(01)}, \tau^{(02)}, \tau^{(10)}, \tau^{(11)}, \right. \nonumber \\
&&& \qquad \left. \tau^{(12)}, \tau^{(23)}, \tau^{(30)}, \tau^{(31)}, \tau^{(32)} \right\} \nonumber \\
(r_{1},r_{2},r_{3},r_{4}) & = (1/3,1/3,+1,+1) & \tau^{(\alpha\beta)} & \in \left\{ \tau^{(03)}, \tau^{(13)}, \tau^{(21)}, \tau^{(22)} \right\}
\end{align}

\subsection{Summary of expressions}

Imposing the $(g_{9})$ and $(h)$ conditions in each case, then in terms of the Hermitian forms,
\begin{align}
H_{\pm}^{(\alpha\beta)} \equiv \psi^{\dagger}\tau^{(\alpha\beta)}\psi \pm \xi^{\dagger}\tau^{(\alpha\beta)}\xi
\end{align}
we have the following generic relations, valid for arbitrary values of all the supergravity fields,
\begin{align}\label{eq:Hermitianmetric} 
f_{2}^{-1}H_{+}^{(00)} & = C_{2}^{(00)} & & \nonumber \\
f_{2}^{1/3}H_{-}^{(\alpha\beta)} & = C_{2}^{(\alpha\beta)} & \tau^{(\alpha\beta)} & \in \left\{ \tau^{(10)}, \tau^{(20)} \right\} \nonumber \\
\nonumber \\
f_{5}^{-1}H_{-}^{(\alpha\beta)} & = C_{5}^{(\alpha\beta)} & \tau^{(\alpha\beta)} & \in \left\{ \tau^{(00)}, \tau^{(10)}, \tau^{(20)}, \tau^{(33)} \right\} \nonumber \\
f_{5}^{3}H_{+}^{(\alpha\beta)} & = C_{5}^{(\alpha\beta)} & \tau^{(\alpha\beta)} & \in \left\{ \tau^{(23)}, \tau^{(30)} \right\} \nonumber \\
\nonumber \\
f_{1}^{-1}H_{-}^{(\alpha\beta)} & = C_{1}^{(\alpha\beta)} & \tau^{(\alpha\beta)} & \in \left\{ \tau^{(00)}, \tau^{(10)}, \tau^{(23)}, \tau^{(30)} \right\}
\end{align}
Since both $f_{2}$ and $\psi^{\dagger}\psi + \xi^{\dagger}\xi$ must be positive, we rescale $\xi$ and $\psi$ by a real constant, so that,
\begin{align}
H_{+}^{(00)} = f_{2}
\end{align}

\section{Vanishing Hermitian forms}\label{sec:vanishinghermforms} 

We can use the reality properties of various combinations of the BPS equations to show that certain Hermitian forms vanish automatically. We consider the following Hermitian forms,
\begin{align}
H_{\pm}^{(\alpha\beta)} & \equiv \psi^{\dagger}\tau^{(\alpha\beta)}\psi \pm \xi^{\dagger}\tau^{(\alpha\beta)}\xi \nonumber \\
H_{g \pm}^{(\alpha\beta)} & \equiv g_{9} \psi^{\dagger}\tau^{(\alpha\beta)}\xi \pm g_{9}^{*}\xi^{\dagger}\tau^{(\alpha\beta)}\psi \nonumber \\
H_{h \pm}^{(\alpha\beta)} & \equiv h\psi^{\dagger}\tau^{(\alpha\beta)}\xi \pm h^{*}\xi^{\dagger}\tau^{(\alpha\beta)}\psi
\end{align}
where $H_{\pm}^{(\alpha\beta)}$, $H_{g +}^{(\alpha\beta)}$, $H_{h +}^{(\alpha\beta)}$ are real, while $H_{g -}^{(\alpha\beta)}$, $H_{h -}^{(\alpha\beta)}$ are purely imaginary. In the following sections, we consider three particular combinations. Then in Section (\ref{sec:Hermitianrelations}), separating out the real and imaginary parts yields the full sets of vanishing and non-trivial Hermitian relations.

\subsection{First set of Hermitian relations}\label{sec:firstHermitianrelations} 

We consider the linear combination $(m) + 2(i) + (9)$ of the BPS equations (\ref{eq:reducedGads}). Note that all the terms containing $f$, $g_{z}$, $g_{\bar{z}}$, $h$, $h^{*}$ are cancelled. Multiplying the first equation by $\xi^{t}\tau^{(\alpha\beta)}$, the second by $-\psi^{t}\tau^{(\alpha\beta) t}$, then adding them and taking the transpose, we obtain,
\begin{align}
0 & = \psi^{\dagger}\left( -\frac{i}{f_{2}}\tau^{(22)} + \frac{2i}{f_{5}}\tau^{(11)} - \frac{\nu}{f_{1}}\tau^{(01)} \right)\tau^{(\alpha\beta)} \psi - \frac{1}{2}g_{9}\psi^{\dagger}\tau^{(02)}\tau^{(\alpha\beta)^{t}}\xi \nonumber \\
& \quad + \xi^{\dagger}\left( -\frac{i}{f_{2}}\tau^{(22)} - \frac{2i}{f_{5}}\tau^{(11)} + \frac{\nu}{f_{1}}\tau^{(01)} \right)\tau^{(\alpha\beta)^{t}} \xi + \frac{1}{2}g_{9}^{*}\xi^{\dagger}\tau^{(02)}\tau^{(\alpha\beta)}\psi
\end{align}

\subsection{Second set of Hermitian relations}\label{sec:secondHermitianrelations} 

We eliminate the $D_{z}f_{i}$, $g_{z}$, $g_{\bar{z}}^{*}$ terms in each set of equations $(m)$, $(i)$, or $(9)$. We calculate only the $(m)$ and $(i)$ equations, since the relations for the $(9)$ equation can be obtained from a linear combination of the $(m)$ and $(i)$ equations, together with the first set of relations.

For each pair of the $f_{2}$ and $f_{5}$ equations in the BPS equations (\ref{eq:reducedGads}), we multiply the first by $\xi^{t}\tau^{(\alpha\beta)}$ and the second by $\psi^{t}\tau^{(\alpha\beta)}$. The $g_{z}$, $g_{\bar{z}}^{*}$ terms then vanish automatically if $\tau^{(\alpha\beta)^{t}} = -\tau^{(\alpha\beta)}$. Adding both to cancel the $D_{z}f_{i}$ terms, then taking the transpose, we have,
\begin{align}
& (m): & 0 & = \psi^{\dagger}\left( \frac{i}{f_{2}}\tau^{(22)} + f\tau^{(12)} \right)\tau^{(\alpha\beta)}\psi  - \frac{3}{8}g_{9}\psi^{\dagger}\tau^{(02)}\tau^{(\alpha\beta)}\xi - \frac{1}{8}h\psi^{\dagger}\tau^{(12)}\tau^{(\alpha\beta)}\xi \nonumber \\
&&& \quad + \xi^{\dagger}\left( -\frac{i}{f_{2}}\tau^{(22)} - f\tau^{(12)} \right)\tau^{(\alpha\beta)}\xi - \frac{3}{8}g_{9}^{*}\xi^{\dagger}\tau^{(02)}\tau^{(\alpha\beta)}\psi + \frac{1}{8}h^{*}\xi^{\dagger}\tau^{(12)}\tau^{(\alpha\beta)}\psi \nonumber \\
& (i): & 0 & = \psi^{\dagger}\left( -\frac{i}{f_{5}}\tau^{(11)} - f\tau^{(12)} \right)\tau^{(\alpha\beta)}\psi + \frac{1}{8}g_{9}\psi^{\dagger}\tau^{(02)}\tau^{(\alpha\beta)}\xi - \frac{1}{8}h\psi^{\dagger}\tau^{(12)}\tau^{(\alpha\beta)}\xi \nonumber \\
&&& \quad + \xi^{\dagger}\left( -\frac{i}{f_{5}}\tau^{(11)} + f\tau^{(12)} \right)\tau^{(\alpha\beta)}\xi + \frac{1}{8}g_{9}^{*}\xi^{\dagger}\tau^{(02)}\tau^{(\alpha\beta)}\psi + \frac{1}{8}h^{*}\xi^{\dagger}\tau^{(12)}\tau^{(\alpha\beta)}\psi
\end{align}

\subsection{Third set of Hermitian relations}\label{sec:thirdHermitianrelations} 

Finally, we consider the combination $(i) - (9)$. We multiply the first equation by $\xi^{t}\tau^{(\alpha\beta)}$ and the second by $\psi^{t}\tau^{(\alpha\beta)}$, with $\tau^{(\alpha\beta)^{t}} = +\tau^{(\alpha\beta)}$. Taking the difference and then the transpose,
\begin{align}
0 & = \psi^{\dagger}\left( \frac{i}{f_{5}}\tau^{(11)} + \frac{\nu}{f_{1}}\tau^{(01)} + 2f\tau^{(12)} \right)\tau^{(\alpha\beta)}\psi + \psi^{\dagger}\left( \frac{1}{2}g_{9}\tau^{(02)} - \frac{1}{2}h\tau^{(12)} \right)\tau^{(\alpha\beta)}\xi \nonumber \\
& \quad + \xi^{\dagger}\left( -\frac{i}{f_{5}}\tau^{(11)} - \frac{\nu}{f_{1}}\tau^{(01)} + 2f\tau^{(12)} \right)\tau^{(\alpha\beta)}\xi + \xi^{\dagger}\left( -\frac{1}{2}g_{9}^{*}\tau^{(02)} - \frac{1}{2}h^{*}\tau^{(12)} \right)\tau^{(\alpha\beta)}\psi
\end{align}

\subsection{Summary of all Hermitian relations}\label{sec:Hermitianrelations} 

The full set of vanishing Hermitian relations is given by,
\begin{align}\label{eq:vanishingforms} 
H_{+}^{\alpha\beta} & = 0 & (\alpha\beta) & \in \left\{ (03), (11), (12), (13), (23), (33) \right\} \nonumber \\
H_{-}^{\alpha\beta} & = 0 & (\alpha\beta) & \in \left\{ (00), (01), (02), (10), (20), (21), (22), (30), (31), (32) \right\} \nonumber \\
\nonumber \\
H_{g+}^{\alpha\beta} & = 0 & (\alpha\beta) & \in \left\{ (00), (10), (23), (30) \right\} \nonumber \\
H_{g-}^{\alpha\beta} & = 0 & (\alpha\beta) & \in \left\{ (03), (11), (12), (13), (31), (32), (33) \right\} \nonumber \\
\nonumber \\
H_{h+}^{\alpha\beta} & = 0 & (\alpha\beta) & \in \left\{ (03), (11), (12), (13), (23) \right\} \nonumber \\
H_{h-}^{\alpha\beta} & = 0 & (\alpha\beta) & \in \left\{ (00), (10), (20), (33) \right\}
\end{align}
The remaining non-trivial Hermitian relations are as follows. We have the first set,
\begin{align}\label{eq:nontrivialfirst} 
& (00) & & \frac{1}{f_{2}}H_{+}^{(22)} - \frac{2}{f_{5}}H_{-}^{(11)} - \frac{i}{2}H_{g-}^{(02)} = 0 \nonumber \\
& (03) & & \frac{1}{f_{2}}H_{+}^{(21)} + \frac{2}{f_{5}}H_{-}^{(12)} - i\frac{1}{2}H_{g-}^{(01)} = 0 \nonumber \\
& (10) & & \frac{1}{f_{2}}H_{+}^{(32)} + \frac{\nu}{f_{1}}H_{-}^{(11)} = 0 \nonumber \\
& (13) & & \frac{1}{f_{2}}H_{+}^{(31)} - \frac{\nu}{f_{1}}H_{-}^{(12)} = 0 \nonumber \\
& (20) & & \frac{2}{f_{5}}H_{+}^{(31)} + \frac{\nu}{f_{1}}H_{+}^{(21)} - \frac{1}{2}H_{g+}^{(22)} = 0 \nonumber \\
& (21) & & \frac{1}{f_{2}}H_{-}^{(03)} + \frac{2}{f_{5}}H_{+}^{(30)} + \frac{\nu}{f_{1}}H_{+}^{(20)} = 0 \nonumber \\
& (22) & & \frac{1}{f_{2}}H_{+}^{(00)} + \frac{2}{f_{5}}H_{-}^{(33)} + \frac{\nu}{f_{1}}H_{-}^{(23)} - \frac{i}{2}H_{g-}^{(20)} = 0 \nonumber \\
& (23) & & \frac{2}{f_{5}}H_{+}^{(32)} + \frac{\nu}{f_{1}}H_{+}^{(22)} + \frac{1}{2}H_{g+}^{(21)} = 0
\end{align}
the second set,
\begin{align}\label{eq:nontrivialsecond} 
& (20) & & 3 H_{g+}^{(22)} + i H_{h-}^{(32)} = 0 & & \frac{1}{f_{5}}H_{+}^{(31)} + \frac{1}{8}H_{g+}^{(22)} - i\frac{1}{8}H_{h-}^{(32)} = 0 \nonumber \\
& (21) & & \frac{1}{f_{2}}H_{-}^{(03)} + f H_{-}^{(33)} = 0 & & \frac{1}{f_{5}}H_{+}^{(30)} - f H_{-}^{(33)} = 0 \nonumber \\
& (23) & & 3 H_{g+}^{(21)} + i H_{h-}^{(31)} = 0 & & \frac{1}{f_{5}}H_{+}^{(32)} - \frac{1}{8}H_{g+}^{(21)} + \frac{i}{8}H_{h-}^{(31)} = 0
\end{align}
and finally the third set,
\begin{align}\label{eq:nontrivialthird} 
& (00) & & \frac{1}{f_{5}}H_{-}^{(11)} - i\frac{1}{2}H_{g-}^{(02)} = 0 \nonumber \\
& (03) & & \frac{1}{f_{5}}H_{-}^{(12)} + i\frac{1}{2}H_{g-}^{(01)} = 0 \nonumber \\
& (10) & & \frac{\nu}{f_{1}}H_{-}^{(11)} + 2f H_{+}^{(02)} - \frac{1}{2}H_{h+}^{(02)} = 0 \nonumber \\
& (13) & & \frac{\nu}{f_{1}}H_{-}^{(12)} - 2f H_{+}^{(01)} + \frac{1}{2}H_{h+}^{(01)}  = 0 \nonumber \\
&(22) && \frac{1}{f_{5}}H_{-}^{(33)} - \frac{\nu}{f_{1}}H_{-}^{(23)} - 2f H_{+}^{(30)} + \frac{i}{2}H_{g-}^{(20)} + \frac{1}{2}H_{h+}^{(30)} = 0 \nonumber \\
& (30) & & 2f H_{+}^{(22)} - \frac{1}{2}H_{h+}^{(22)} = 0 \nonumber \\
& (33) & & 2f H_{+}^{(21)} - \frac{1}{2}H_{h+}^{(21)} = 0
\end{align}

\subsection{Implications for the metric factors}

Together with (\ref{eq:Hermitianmetric}), the above relations imply the vanishing of the following constants,
\begin{align}
0 & = C_{2}^{(10)} = C_{2}^{(20)} \nonumber \\
0 & = C_{5}^{(00)} = C_{5}^{(10)} = C_{5}^{(20)} = C_{5}^{(23)} \nonumber \\
0 & = C_{1}^{(00)} = C_{1}^{(10)} = C_{1}^{(30)}
\end{align}
which leaves the following non-vanishing Hermitian forms,
\begin{align}
f_{2}^{-1}H_{+}^{(00)} & = 1 \nonumber \\
f_{5}^{-1}H_{-}^{(33)} & = C_{5}^{(33)} \nonumber \\
f_{5}^{3}H_{+}^{(30)} & = C_{5}^{(30)} \nonumber \\
f_{1}^{-1}H_{-}^{(23)} & = C_{1}^{(23)}
\end{align}
where we have used the normalization $C_{2}^{(00)} = 1$.

\section{General solutions to the reduced BPS equations}\label{sec:gensolnsBPS} 

In this section, we use the vanishing Hermitian forms to solve the reduced BPS equations. We follow the same procedure and reach the same conclusion as in \cite{DHoker:2010gus}, namely that the only solution to the reduced BPS equations is the maximally supersymmetric solution $AdS_{5}\times S^{5}$.

\subsection{Solving the Hermitian relations $H_{\pm}^{(\alpha\beta)} = 0$}

Grouping the vanishing Hermitian relations $H_{\pm}^{(\alpha\beta)} = 0$ from (\ref{eq:vanishingforms}) into four sets, we obtain the following relations between the spinor components for $\eta_{1} = \pm$ and $\eta_{2} = \pm$ independently,
\begin{align}\label{eq:groups} 
0 & = H_{-}^{(00)} = H_{-}^{(30)} = H_{+}^{(03)} = H_{+}^{(33)} & & \implies & \xi_{\eta_{1},\eta_{2}}^{*}\xi_{\eta_{1},\eta_{2}} - \psi_{\eta_{1},-\eta_{2}}^{*}\psi_{\eta_{1},-\eta_{2}} & = 0 \nonumber \\
0 & = H_{-}^{(10)} = H_{-}^{(20)} = H_{+}^{(13)} = H_{+}^{(23)} & & \implies & \xi_{\eta_{1},\eta_{2}}^{*}\xi_{-\eta_{1},\eta_{2}} - \psi_{\eta_{1},-\eta_{2}}^{*}\psi_{-\eta_{1},-\eta_{2}} & = 0 \nonumber \\
0 & = H_{-}^{(01)} = H_{-}^{(31)} = H_{-}^{(02)} = H_{-}^{(32)} & & \implies & \xi_{\eta_{1},\eta_{2}}^{*}\xi_{\eta_{1},-\eta_{2}} - \psi_{\eta_{1},\eta_{2}}^{*}\psi_{\eta_{1},-\eta_{2}} & = 0 \nonumber \\
0 & = H_{+}^{(11)} = H_{-}^{(21)} = H_{+}^{(12)} = H_{-}^{(22)} & & \implies & \xi_{\eta_{1},\eta_{2}}^{*}\xi_{-\eta_{1},-\eta_{2}} + \psi_{-\eta_{1},\eta_{2}}^{*}\psi_{\eta_{1},-\eta_{2}} & = 0
\end{align}
When the $\psi_{\eta_{1},\eta_{2}}$ are all generic and non-vanishing, the solutions to (\ref{eq:groups}) are of the form,
\begin{align}\label{eq:firsttype} 
\psi_{++} & = r_{++}e^{i\Lambda + i\Phi} & \xi_{++} & = e^{i\theta_{1}}\psi_{+-} = r_{+-}e^{i\Lambda' + i\Phi} \nonumber \\
\psi_{+-} & = r_{+-}e^{i\Lambda - i\Phi} & \xi_{+-} & = e^{i\theta_{2}}\psi_{++} = r_{++}e^{i\Lambda' - i\Phi} \nonumber \\
\psi_{-+} & = r_{-+}e^{i\Lambda + i\Phi + i \pi/2} & \xi_{-+} & = e^{i\theta_{1}}\psi_{--} = r_{--}e^{i\Lambda' + i\Phi + i \pi/2} \nonumber \\
\psi_{--} & = r_{--}e^{i\Lambda - i\Phi + i \pi/2} & \xi_{--} & = e^{i\theta_{2}}\psi_{-+} = r_{-+}e^{i\Lambda' - i\Phi + i \pi/2}
\end{align}
parametrized in terms of 4 real functions $r_{\eta_{1},\eta_{2}}$ plus the angles $\theta_{1} = 2\Phi + 2\Phi'$, $\theta_{2} = -2\Phi + 2\Phi'$, $\Lambda' = 2\Phi' + \Lambda$, and $\Lambda$ arbitrary. 
The case where one component $\psi_{\eta_{1},\eta_{2}} = 0$ can be viewed as the limit in which $r_{\eta_{1},\eta_{2}} = 0$. The only exception is when $\psi_{\eta_{1},\eta_{2}} = \psi_{-\eta_{1},\eta_{2}} = 0$ and $\psi_{\eta_{1},-\eta_{2}} = \psi_{-\eta_{1},-\eta_{2}} \neq 0$. We consider the case $\psi_{+-} = \psi_{--} = 0$ and $\psi_{++}, \psi_{-+} \neq 0$, which may be parametrized by four real fuctions $r_{++}\equiv r_{1},r_{-+}\equiv r_{3},\Lambda_{1},\Lambda_{3}$, plus an angle $\theta$, as follows,
\begin{align}\label{eq:secondtype} 
\psi_{++} & = r_{1}e^{i\Lambda_{1}} & \xi_{+-} & = e^{i\theta}\psi_{++} = r_{1}e^{i(\Lambda_{1} + \theta)} \nonumber \\
\psi_{-+} & = r_{3}e^{i\Lambda_{3}} & \xi_{--} & = e^{i\theta}\psi_{-+} = r_{3}e^{i(\Lambda_{3} + \theta)}
\end{align}
The solutions (\ref{eq:firsttype}) - which we will refer to as the ``first type'' of solutions - 
reproduce all the relations $H_{\pm}^{(\alpha\beta)} = 0$ in (\ref{eq:vanishingforms}), as well as two additional relations not listed in Section (\ref{sec:Hermitianrelations}),
\begin{align}
H_{+}^{(10)} & = 0 & H_{-}^{(13)} & = 0
\end{align}
The solutions (\ref{eq:secondtype}) - which we will refer to as the ``second type'' of solutions - 
reproduce all the relations $H_{\pm}^{(\alpha\beta)} = 0$ in (\ref{eq:vanishingforms}), as well as the following additional vanishing conditions,
\begin{align}
0 = H_{+}^{(01)} = H_{+}^{(02)} = H_{+}^{(21)} = H_{+}^{(22)} = H_{+}^{(31)} = H_{+}^{(32)} = H_{-}^{(11)} = H_{-}^{(12)}
\end{align}

\subsection{Solving the Hermitian relations $H_{g\pm}^{(\alpha\beta)}$, $H_{h\pm}^{(\alpha\beta)}$}

Next, we use the solutions of the previous section to obtain conditions from the remaining vanishing Hermitian forms. It is then straightforward but tedious to show that the only non-trivial solutions are those with $g_{9} = 0$. The calculation parallels the one in \cite{DHoker:2010gus}, so we will summarize the results while highlighting any notable differences. We define the quantities,
\begin{align}
r & = 
\begin{pmatrix}
r_{++} \\
r_{+-} \\
r_{-+} \\
r_{--}
\end{pmatrix}
\equiv
\begin{pmatrix}
r_{1} \\
r_{2} \\
r_{3} \\
r_{4}
\end{pmatrix}
&
\begin{Bmatrix}
\mathcal{G}_{\pm} \\
\\
\mathcal{H}_{\pm}
\end{Bmatrix}
& \equiv
\begin{Bmatrix}
g_{9}e^{i(\Lambda' - \Lambda)} \pm g_{9}^{*}e^{-i(\Lambda' - \Lambda)} \\
\\
h e^{i(\Lambda' - \Lambda)} \pm h^{*}e^{-i(\Lambda' - \Lambda)}
\end{Bmatrix}
\end{align}

\subsubsection{First type of solution}

For the first type of solutions (\ref{eq:firsttype}), the following Hermitian forms vanish automatically,
\begin{align}
H_{g\pm}^{(\alpha\beta)} & = H_{h\pm}^{(\alpha\beta)} = 0 & (\alpha\beta) & \in \left\{ (03), (10), (11), (12), (23), (33) \right\}
\end{align}
The remaining vanishing Hermitian forms (\ref{eq:vanishingforms}) yield two sets of conditions for $H_{g\pm}^{(\alpha\beta)} = 0$:
\begin{align}\label{eq:g9condition1} 
\mathcal{G}_{+}r^{t}\tau^{(\gamma\delta)}r & = 0 & (\gamma\delta) & \in \left\{ (01), (22), (31) \right\}
\end{align}
\begin{align}\label{eq:g9condition2} 
\mathcal{M}_{g}
\begin{pmatrix}
r^{t}\tau^{(30)}r \\
r^{t}\tau^{(33)}r
\end{pmatrix}
\equiv
\begin{pmatrix}
\cos\left( 2\Phi \right)\mathcal{G}_{-} & -i\sin\left( 2\Phi \right)\mathcal{G}_{+} \\
-\sin\left( 2\Phi \right)\mathcal{G}_{-} & -i\cos\left( 2\Phi \right)\mathcal{G}_{+}
\end{pmatrix}
\begin{pmatrix}
r^{t}\tau^{(30)}r \\
r^{t}\tau^{(33)}r
\end{pmatrix}
= 0
\end{align}
From the relations (\ref{eq:vanishingforms}), we also have a set of conditions for the Hermitian forms $H_{h\pm}^{(\alpha\beta)} = 0$:
\begin{align}\label{eq:hcondition1a} 
\mathcal{H}_{-}r^{t}\tau^{(\gamma\delta)}r & = 0 & (\gamma\delta) & \in \left\{ (01), (11), (22) \right\}
\end{align}
From (\ref{eq:g9condition2}) we have $\det \mathcal{M}_{g} = -i \mathcal{G}_{+}\mathcal{G}_{-}$. For trivial solutions corresponding to $\mathcal{G}_{+}\mathcal{G}_{-} \neq 0$, the conditions (\ref{eq:g9condition1}) and (\ref{eq:g9condition2}), together with the $(22)$ relations in (\ref{eq:nontrivialfirst}) and (\ref{eq:nontrivialthird}), imply $H_{+}^{(00)} = 0$ and thus all $r_{\eta_{1},\eta_{2}} = 0$. Non-trivial solutions correspond to $\mathcal{G}_{+}\mathcal{G}_{-} = 0$, and one can show that for either choice $\mathcal{G}_{\pm} = 0$ and $\mathcal{G}_{\mp} \neq 0$, the conditions (\ref{eq:g9condition1}) and (\ref{eq:g9condition2}) plus the non-trivial relations from Section (\ref{sec:Hermitianrelations}), imply that all $r_{\eta_{1},\eta_{2}} = 0$. For example, if $\mathcal{G}_{+} = 0$ and $\mathcal{G}_{-} \neq 0$ then (\ref{eq:g9condition1}) is automatically satisfied, while (\ref{eq:g9condition2}) and the second $(21)$ relation in (\ref{eq:nontrivialsecond}) imply,
\begin{align}
r^{t}\tau^{(30)}r & = 0: & H_{-}^{(33)} = \frac{1}{f f_{5}}H_{+}^{(30)} & = 0 & & \implies & r^{t}\tau^{(33)}r & = 0
\end{align}
The Hermitian forms that vanish under $r^{t}\tau^{(30)}r = r^{t}\tau^{(33)}r = 0$ cause a number of relations in Section (\ref{sec:Hermitianrelations}) to become trivial, which in turn produce conditions that can only be satisfied if all $r_{\eta_{1},\eta_{2}} = 0$. The only remaining possibility is $g_{9} = 0$, which yields extra vanishing forms:
\begin{align}\label{eq:g9zerocondition} 
& 0 = H_{+}^{(21)} = H_{+}^{(22)} = H_{+}^{(31)} = H_{+}^{(32)} = H_{-}^{(11)} = H_{-}^{(12)} \nonumber \\
& 0 = H_{h+}^{(21)} = H_{h+}^{(22)} = H_{h-}^{(31)} = H_{h-}^{(32)}
\end{align}
The top line of (\ref{eq:g9zerocondition}) plus the original vanishing Hermitian forms imply the conditions,
\begin{align}
r_{1}r_{4} & = r_{2}r_{3} = 0 & r_{1}r_{2} - r_{3}r_{4} & = 0
\end{align}
Without loss of generality, we choose $r_{4} = 0$, so that either $r_{1} = r_{3} = 0$ or $r_{2} = 0$, and examine the dilatino equation (\ref{eq:reducedD}). If $r_{1}, r_{3} \neq 0$ and $r_{2} = r_{4} = 0$, then 
we must have,
\begin{align}\label{eq:g9spinorconds} 
p_{z} & = p_{\bar{z}} = 0 & |h|^{2} - |g_{z}|^{2} & = 0 & |h|^{2} - |g_{\bar{z}}|^{2} & = 0
\end{align}
for non-vanishing spinor solutions. But in order to have non-trivial solutions while satisfying both the original conditions (\ref{eq:hcondition1a}) and the bottom line of (\ref{eq:g9zerocondition}), we must set $h = 0$ so that,
\begin{align}\label{eq:g9spinorcomps} 
p_{z} = p_{\bar{z}} = g_{z} = g_{\bar{z}} = h = 0
\end{align}
On the other hand, if we take $r_{1} = r_{3} = r_{4} = 0$ and $r_{3} \neq 0$, then this result is automatic.

\subsubsection{Second type of solution}

For the second type of solutions (\ref{eq:secondtype}), the following Hermitian forms vanish automatically,
\begin{align}
H_{g\pm}^{(\alpha\beta)} & = H_{h\pm}^{(\alpha\beta)} = 0 & (\alpha\beta) & \in \left\{ (00), (03), (10), (13), (20), (23), (30), (33) \right\}
\end{align}
and the extra forms $H_{h+}^{(30)} = H_{g-}^{(20)} = 0$ modify the $(22)$ relations in (\ref{eq:nontrivialfirst}) and (\ref{eq:nontrivialthird}) as follows,
\begin{align}\label{eq:2ndGforms} 
\frac{1}{f_{2}}H_{+}^{(00)} + \frac{2}{f_{5}}H_{-}^{(33)} + \frac{\nu}{f_{1}}H_{-}^{(23)} & = 0 & \frac{1}{f_{5}}H_{-}^{(33)} - \frac{\nu}{f_{1}}H_{-}^{(23)} - 2f H_{+}^{(30)} & = 0
\end{align}
From the remaining cases of $H_{g\pm}^{(\alpha\beta)} = 0$ and $H_{h\pm}^{(\alpha\beta)} = 0$ in (\ref{eq:vanishingforms}), we obtain the conditions,
\begin{align}\label{eq:2ndGcondition} 
\left[ e^{i(\Lambda_{1}-\Lambda_{3})} + e^{-i(\Lambda_{1}-\Lambda_{3})} \right] \left( e^{i\theta}g_{9} \pm e^{-i\theta}g_{9}^{*} \right) r_{1}r_{3} & = 0 \nonumber \\
\left( e^{i\theta}g_{9} \pm e^{-i\theta}g_{9}^{*} \right)\left( r_{1}^{2} - r_{3}^{2} \right) & = 0
\end{align}
\begin{align}\label{eq:2ndHcondition} 
\left[ e^{i(\Lambda_{1} - \Lambda_{3})} + e^{-i(\Lambda_{1} - \Lambda_{3})} \right] \left( e^{i\theta}h \pm e^{-i\theta}h^{*} \right)r_{1}r_{3} = 0
\end{align}
For non-trivial solutions with $g_{9} \neq 0$, we must have $\textrm{Re}\left[ e^{i(\Lambda_{1}-\Lambda_{3})} \right] = 0$ and $r_{1}^{2}=r_{3}^{2}$. Under this choice, the $(22)$ relations of (\ref{eq:nontrivialfirst}) and (\ref{eq:nontrivialthird}) reduce to $H_{+}^{(00)} = 0$ and thus all $r_{\eta_{1},\eta_{2}} = 0$. So we again must have $g_{9} = 0$, which then yields the additional vanishing Hermitian forms,
\begin{align}\label{eq:g9zerocondition2} 
0 = H_{h+}^{(21)} = H_{h+}^{(22)} = H_{h-}^{(31)} = H_{h-}^{(32)}
\end{align}
Examining the dilatino equation (\ref{eq:reducedD}), we find the same constraints as (\ref{eq:g9spinorconds}) on the supergravity fields. The extra forms in (\ref{eq:2ndGforms}) together with (\ref{eq:g9zerocondition2}) impose the following conditions:
\begin{align}
\Lambda_{\eta_{1}}\Theta_{\eta_{2}}r_{1}r_{3} & = 0 & \Theta_{\eta_{3}}\left( r_{1}^{2} - r_{3}^{2} \right) & = 0
\end{align}
where the $\eta_{i} = \pm$ for $i=1,2,3$ independently, and we have defined the quantities,
\begin{align}
\Lambda_{\eta_{1}} & = e^{i(\Lambda_{1} - \Lambda_{3})} + \eta_{1} e^{-i(\Lambda_{1} - \Lambda_{3})} & \Theta_{\eta_{2}} & = e^{i\theta}h + \eta_{2} e^{-i\theta}h^{*}
\end{align}
An analysis similar to the one used for the first type of solution again yields the result (\ref{eq:g9spinorcomps}).

\subsection{Vanishing $G$ implies the $AdS_{5}\times S^{5}$ solution}

When $G=0$, we have $g_{z} = g_{\bar{z}} = g_{9} = h = 0$. For half-BPS solutions, $\psi$ and $\xi$ cannot both vanish, and the reduced dilatino equation (\ref{eq:reducedGsigma}) implies $p_{z} = p_{\bar{z}} = 0$. By the Bianchi identities (\ref{eq:Bianchi}), $P = 0$ implies $dQ = 0$, and we use the $U(1)_{q}$ gauge symmetry to set $Q = 0$. Therefore, the requirements (\ref{eq:g9spinorcomps}) can be obtained directly by imposing the vanishing of $G$.

\subsubsection{Using the discrete symmetries}

The generators $\tau^{(033)}, \tau^{(030)}$ (\ref{eq:discretesymm}) and $\mathcal{K}$ (\ref{eq:CCsymm}) may be simultaneously diagonalized as follows,
\begin{align}
\tau^{(033)}\zeta & = \nu\zeta & \tau^{(030)}\zeta & = \gamma\zeta & \mathcal{K}\zeta & = \mu\zeta
\end{align}
where $\nu, \gamma, \mu$ take on the values $\pm 1$ independently. The $\tau^{(033)}$ projection was used to obtain the chiral form of the reduced BPS equation. We define the projections of $\tau^{(030)}$ and $\mathcal{K}$ as,
\begin{align}
\tau^{(030)}: & & \tau^{(03)}\psi & = \gamma\psi & \tau^{(03)}\xi & = -\gamma\xi \nonumber \\
\mathcal{K}: & & \xi & = \tau^{(02)}\psi & \mu & \equiv -\nu e^{i\theta}
\end{align}
using the $U(1)_{q}$ gauge symmetry to fix the sign of the $\mathcal{K}$ projection. For $\gamma = +1$, we take,
\begin{align}
\psi & =
\begin{pmatrix}
\psi_{+} \\
\psi_{-}
\end{pmatrix}
=
\begin{pmatrix}
\psi_{++} \\
\psi_{-+}
\end{pmatrix}
& \xi & =
\begin{pmatrix}
\xi_{+} \\
\xi_{-}
\end{pmatrix}
=
\begin{pmatrix}
\xi_{+-} \\
\xi_{--}
\end{pmatrix}
= -i
\begin{pmatrix}
\psi_{++} \\
\psi_{-+}
\end{pmatrix}
\end{align}
to be two-component spinors with $\eta_{2}$ fixed. The remaining reduced BPS equations are then,
\begin{align}\label{eq:BPSeqnsGzero} 
(m) & & & \pm\frac{1}{f_{2}}\psi_{\mp}^{*} + \frac{D_{z}f_{2}}{f_{2}}\psi_{\pm} - f\psi_{\mp}^{*} = 0 \nonumber \\
(i) & & & -\frac{1}{f_{5}}\psi_{\mp}^{*} + \frac{D_{z}f_{5}}{f_{5}}\psi_{\pm} + f\psi_{\mp}^{*} = 0 \nonumber \\
(9) & & & -\frac{i\nu}{f_{1}}\psi_{\pm}^{*} + \frac{D_{z}f_{1}}{f_{1}}\psi_{\pm} - f\psi_{\mp}^{*} = 0 \nonumber \\
(-) & & & \left( D_{\bar{z}} - \frac{i}{2}\hat{\omega}_{\bar{z}} \right)\psi_{\pm} = 0 \nonumber \\
(+) & & & \left( D_{z} - \frac{i}{2}\hat{\omega}_{z} \right)\psi_{\pm} - f\psi_{\mp}^{*} = 0
\end{align}

\subsubsection{Generic solutions when $G = 0$}

Using $\hat{\omega}_{z}=i(\partial_{z}\rho)/\rho^{2}$ and $D_{z} = \rho^{-1}\partial_{z}$, the solution to the $(-)$ equation of (\ref{eq:BPSeqnsGzero}) is given by,
\begin{align}\label{eq:PSIsolns} 
\psi_{+} & = \sqrt{\rho}\alpha & \psi_{-} & = \sqrt{\rho}\beta & & \partial_{\bar{z}}\alpha = \partial_{\bar{z}}\beta = 0
\end{align}
Employing the same strategy as \cite{DHoker:2010gus}, the $(\pm)$ equations are used in combination with the $(m)$, $(i)$, $(9)$ equations of (\ref{eq:BPSeqnsGzero}) to obtains solutions in terms of $\psi_{\pm}$ for the metric factors,
\begin{align}
f_{2} & = |\psi_{+}|^{2} + |\psi_{-}|^{2} , & f_{5} & = |\psi_{+}|^{2} - |\psi_{-}|^{2} , & f_{1} & = c_{1}\left( \psi_{+}^{*}\psi_{-} - \psi_{-}^{*}\psi_{+} \right)
\end{align}
where $D_{z}f_{5} = 0$ and we set $f_{5} \equiv 1$. We can rewrite the $(\pm)$ equations (\ref{eq:BPSeqnsGzero}) involving $f_{5}$ as,
\begin{align}\label{eq:MMbareqn} 
\left| \frac{D_{z}f_{5}}{f_{5}} \right|^{2}\psi_{+}\psi_{-}^{*} - \left( \frac{1}{f_{5}} - f \right)^{2}\psi_{+}\psi_{-}^{*} = 0
\end{align}
If $\psi_{+}\psi_{-}^{*} \neq 0$, then $D_{z}f_{5} = 0$ implies that $f f_{5} = 1$. If either $\psi_{+} = 0$ or $\psi_{-}^{*} = 0$, the $(i)$ equations of (\ref{eq:BPSeqnsGzero}) also imply that $f f_{5} = 1$. The $(+)$ equations of (\ref{eq:BPSeqnsGzero}) then reduce to,
\begin{align}
\beta\partial_{z}\alpha - \alpha\partial_{z}\beta + 1 = 0
\end{align}
The solution is given, in terms of an arbitrary holomorphic function $A(z)$, by the expressions,
\begin{align}
\alpha(z) & = \frac{1}{\sqrt{\partial_{z}A(z)}} & \beta(z) & = \frac{A(z)}{\sqrt{\partial_{z}A(z)}}
\end{align}

\subsubsection{Solution of $AdS_{5}\times S^{5}$}

Choosing $A(z) = -e^{-2z}$, with $c_{1} = i$ so that $f_{1}$ is real, the 10-dimensional metric becomes,
\begin{align}
ds^{2} = \left( \coth x_{7} \right)^{2} ds_{AdS_{2}}^{2} + ds_{S^{5}}^{2} + \frac{dx_{7}^{2} + dx_{8}^{2}}{(\sinh x_{7})^{2}} + \frac{\sin^{2}x_{8} dx_{9}^{2}}{(\sinh x_{7})^{2}}
\end{align}
where $z = (x_{7} + i x_{8})/2$. Performing the following transformation on the $x_{7}$ coordinate,
\begin{align}
e^{x_{7}} = \tanh \left( \frac{\theta}{2} \right)
\end{align}
we recover the $AdS_{5}\times S^{5}$ metric in the standard form,
\begin{align}
ds^{2} & = \cosh^{2}\theta \, ds_{AdS_{2}}^{2} + ds_{S^{5}}^{2} + \sinh^{2}\theta\left( dx_{7}^{2} + dx_{8}^{2} \right) + \sinh^{2}\theta \sin^{2}x_{8} dx_{9}^{2} \nonumber \\
& = \cosh^{2}\theta \, ds_{AdS_{2}}^{2} + ds_{S^{5}}^{2} + d\theta^{2} + \sinh^{2}\theta dx_{8}^{2} + \sinh^{2}\theta \sin^{2}x_{8} dx_{9}^{2} \nonumber \\
& = \left[ d\theta^{2} + \cosh^{2}\theta \, ds_{AdS_{2}}^{2} + \sinh^{2}\theta \, ds_{S^{2}}^{2} \right] + ds_{S^{5}}^{2}
\end{align}
The solution to the spinor $\zeta$ is characterized by the three projections,
\begin{align}
\sigma^{1}\zeta^{*} & = -\zeta & \tau^{(033)}\zeta & = \nu\zeta & \tau^{(030)}\zeta & = \gamma\zeta
\end{align}
With 8 independent Killing spinors $\chi$ in (\ref{eq:complexspinor}) and 4 independent solutions to $\zeta$ of the form,
\begin{align}
\nu & = \eta_{2}\eta_{3} & \gamma & = \eta_{2}: & \zeta_{\pm,\eta_{2},\eta_{3}} & =
\begin{pmatrix} 
\zeta_{\pm} \\
-\bar{\zeta}_{\pm} \\
\end{pmatrix}
\end{align}
we indeed recover 32 supersymmetries for the maximally supersymmetric solution $AdS_{5}\times S^{5}$.

\section{Conclusion}\label{sec:conclusion} 

We have proven that for a spacetime of the form $AdS_{2}\times S^{5}\times S^{1}$ warped over a Riemann surface $\Sigma$, the only solution with at least 16 supersymmetries is just the maximally supersymmetric solution $AdS_{5}\times S^{5}$. As we discussed, this then implies that no supergravity solutions exist for fully back-reacted D7 probe or D7/D3 intersecting branes whose near-horizon limit has the same spacetime structure with corresponding $SO(2,1)\oplus SO(6)\oplus SO(2)$ symmetry. 
Thus the $SU(1,1|4)$-invariant $AdS_{2}$ solutions are \textit{rigid} in exactly the same sense as the two $SU(2,2|2)$-symmetric cases considered in \cite{DHoker:2010gus},
while the case of $SU(1,1|4)\oplus SU(1,1)$ is left for consideration in a future work. 


The procedure employed in this paper involved constructing the most general Ansatz for the bosonic Type IIB supergravity fields that can be realized for the $AdS_{2}\times S^{5}\times S^{1}\times \Sigma$ spacetime geometry. Thus the only restriction on possible sources for e.g. the complex 3-form $G$ or the 5-form field strength $F_{(5)}$ is the $SO(2,1)\oplus SO(6)\oplus SO(2)$ symmetry. For example, in \cite{DHoker:2016ujz, DHoker:2016ysh, DHoker:2017mds, DHoker:2017zwj} and \cite{Corbino:2017tfl, Corbino:2018fwb} the two- and six-form potentials (respectively) were ultimately valuable indicators of brane and string sources for the physically regular solutions. But for the present case, we found that the relations following from the reduction of the BPS equations imply that $G = 0$ while the coefficient of $F_{(5)}$ is constant, thus precluding the existence of non-trivial sources for such half-BPS solutions. Relaxing the condition of having 16 supersymmetries could allow for additional sources, but then one would have to modify the Ansatz for the supersymmetry generators. This in turn would require an entirely different strategy for obtaining solutions, as the use of the Killing spinors of the maximally symmetric subspaces provides a crucial ingredient in the reduction of the BPS equations.


Finally, we observe that for both the present case as well as the two cases in \cite{DHoker:2010gus}, one of the internal factors of the corresponding maximally supersymmetric solution is present in the warped spacetime of the half-BPS solutions, either $AdS_{5}$ or $S^{5}$ for Type IIB supergravity. One can show (as was done in Sec. 2.3 of \cite{DHoker:2010gus}) that in each case the Bianchi identity for $F_{(5)}$ yields the same constraint on the corresponding metric factor and field strength as was obtained by solving the BPS equations. Namely, the condition that the product $f f_{5}^{5}$ is constant, which in each case could only be satisfied if both $f$ and $f_{5}$ are constant, and in turn leaves only the $AdS_{5}\times S^{5}$ solution. An open question is whether such rigidity extends to any half-BPS solution that has a spacetime factor in common with the corresponding maximally supersymmetric solution, for example warped $AdS_{2}\times S^{7}$ solutions to M-theory.

\section*{Acknowledgements}


The author would like to thank Michael Gutperle, Justin Kaidi, and especially Eric D'Hoker for many helpful conversations. He would also like to thank the Mani L. Bhaumik Institute for Theoretical Physics for continued support. Finally, special thanks to Kelly Blumenthal.


\appendix

\section{Clifford algebra basis adapted to the Ansatz}\label{sec:Clifford} 

The Dirac-Clifford algebra is defined by $\{\Gamma^{A},\Gamma^{B} \} = 2\eta^{AB}I_{32}$, where $A, B$ are 10-dimensional frame indices and $\eta^{AB} = \textrm{diag}(- + \cdots +)$. We choose a basis for the Clifford algebra which is well-adapted to the $AdS_{2} \times S^{5} \times \Sigma \times S^{1}$ Ansatz, with the frame labeled as in (\ref{eq:frames}),
\begin{align}
\Gamma^{m} &= \gamma^{m} \otimes I_{4} \otimes I_{2} \otimes \sigma^{1} & m & = 0,1 \nonumber \\
\Gamma^{i} &= I_{2} \otimes \gamma^{i} \otimes I_{2} \otimes \sigma^{3} & i & =  2, 3, 4, 5, 6 \nonumber \\
\Gamma^{a} &= \sigma^{3} \otimes I_{4} \otimes \gamma^{a} \otimes \sigma^{1} & a & = 7, 8 \nonumber \\
\Gamma^{9} & = \sigma^{3} \otimes I_{4} \otimes \gamma^{9} \otimes \sigma^{1} &&
\end{align}
where the lower dimensional Dirac-Clifford algebra is defined as follows,
\begin{align}
\gamma^{0} & = i\sigma^{1} & \gamma^{2} &= \sigma^{1} \otimes I_{2} && \nonumber \\
\gamma^{1} & = \sigma^{2} & \gamma^{3} &= \sigma^{2} \otimes I_{2} && \nonumber \\
&& \gamma^{4} &= \sigma^{3} \otimes \sigma^{1} & \gamma^{7} & = \sigma^{1} \nonumber \\
&&\gamma^{5} &= \sigma^{3} \otimes \sigma^{2} & \gamma^{8} & = \sigma^{2} \nonumber \\
&& \gamma^{6} &= \sigma^{3} \otimes \sigma^{3} & \gamma^{9} & = \sigma^{3}
\end{align}
The chirality matrices on the various components of $AdS_{2} \times S^{5} \times \Sigma \times S^{1}$ are given by,
\begin{align}
\Gamma^{01} & = -\gamma_{(1)} \otimes I_{4} \otimes I_{2} \otimes I_{2} & \gamma_{(1)} & = -\gamma^{0}\gamma^{1} = \sigma^{3} \nonumber \\
\Gamma^{23456} & = -I_{2} \otimes \gamma_{(2)} \otimes I_{2} \otimes \sigma^{3} & \gamma_{(2)} & = -\gamma^{2}\cdots\gamma^{6} = I_{4} \nonumber \\
\Gamma^{78} & = i I_{2} \otimes I_{4} \otimes \gamma_{(3)} \otimes I_{2} & \gamma_{(3)} & = -i\gamma^{7}\gamma^{8} = \sigma^{3}
\end{align}
which yields the following 10-dimensional chirality matrix,
\begin{equation}
\Gamma^{11} = \Gamma^{0123456789} = -I_{2}\otimes I_{4}\otimes I_{2}\otimes \sigma^{2}
\end{equation}
The complex conjugation matrices in each component are defined by,
\begin{align}
(\gamma^{m})^{*} &= - B_{(1)}\gamma^{m} B_{(1)}^{-1} & \left( B_{(1)} \right)^{*} B_{(1)} & = +I_{2} & B_{(1)} & = I_{2} \nonumber \\
(\gamma^{i})^{*} &= + B_{(2)}\gamma^{i}B^{-1}_{(2)} & \left( B_{(2)} \right)^{*} B_{(2)} & = -I_{4} & B_{(2)} & = \sigma^{1} \otimes \sigma^{2} \nonumber \\
(\gamma^{a})^{*} &= - B_{(3)}\gamma^{a}B^{-1}_{(3)} & \left( B_{(3)} \right)^{*} B_{(3)} & = -I_{2} & B_{(3)} & = \sigma^{2} \nonumber \\
(\gamma^{9})^{*} &= +B_{(4)}\gamma^{9}B^{-1}_{(4)} & \left( B_{(4)} \right)^{*} B_{(4)} & = +I_{2} & B_{(4)} & = I_{2}
\end{align}
where in the last column we have also listed the form of these matrices in our particular basis. The 10-dimensional complex conjugation matrix $\mathcal{B}$ satisfies,
\begin{align}
(\Gamma^{M})^{*} &=  \mathcal{B}\Gamma^{M}\mathcal{B}^{-1} & \mathcal{B}^{*} \mathcal{B} & = I_{32} & \{ \mathcal{B},\Gamma^{11} \} & = 0
\end{align}
and in this basis has the following form,
\begin{align}
\mathcal{B} = I_{2} \otimes \sigma^{1} \otimes \sigma^{2} \otimes \sigma^{2} \otimes \sigma^{3}
\end{align}

\section{Deriving the reduced BPS equations}\label{sec:BPS}

In reducing the BPS equations, we will use the following decompositions of $\varepsilon$ and $\mathcal{B}^{-1}\varepsilon^{*}$,
\begin{align}
\varepsilon & = \sum_{\eta_{1},\eta_{2},\eta_{3}}\chi^{\eta_{1},\eta_{2}}\chi^{\eta_{3}}\otimes\zeta_{\eta_{1},\eta_{2},\eta_{3}}\otimes\phi & \mathcal{B}^{-1}\varepsilon^{*} & = \sum_{\eta_{1},\eta_{2},\eta_{3}}\chi^{\eta_{1},\eta_{2}}\chi^{\eta_{3}}\otimes\star\zeta_{\eta_{1},\eta_{2},\eta_{3}}\otimes\phi
\end{align}
where we have used the abbreviations,
\begin{align}\label{eq:stardef} 
\star\zeta_{\eta_{1},\eta_{2},\eta_{3}} & = -i\eta_{2}\sigma^{2}\zeta^{*}_{-\eta_{1},-\eta_{2},-\eta_{3}} & \star\zeta & = \tau^{(121)}\sigma^{2}\zeta^{*}
\end{align}
in $\tau$-matrix notation. The strategy employed here is the same as the one used in Appendix B of \cite{DHoker:2010gus}, therefore we will summarize the derivation while highlighting certain key details.

\subsection{The dilatino equation}

Using the explicit form of the supergravity fields and $\Gamma$-matrices, the dilatino equation is,
\begin{align}
0 = \sum_{\eta_{1},\eta_{2},\eta_{3}}\chi^{\eta_{1},\eta_{2}}\chi^{\eta_{3}}\otimes\left[ \eta_{2}p_{a}\gamma^{a}\sigma^{2}\zeta_{\eta_{1},-\eta_{2},-\eta_{3}}^{*} - \frac{1}{4}g_{\bar{a}}\gamma^{\bar{a}}\zeta_{\eta_{1},\eta_{2},\eta_{3}} + \frac{1}{4}h\zeta_{-\eta_{1},\eta_{2},\eta_{3}} \right]\otimes \phi^{*}
\end{align}
The linear independence of the $\chi^{\eta_{1},\eta_{2}}\chi^{\eta_{3}}$ implies that each coefficient in the square brackets must vanish separately. Rewriting the result in terms of $\tau$-matrix notation, we recover (\ref{eq:dilatino}):
\begin{align}
0 = 4p_{a}\gamma^{a}\sigma^{2}\zeta^{*} + i g_{\bar{a}}\tau^{(021)}\gamma^{\bar{a}}\zeta - i h\tau^{(121)}\zeta
\end{align}

\subsection{The gravitino equation}

The components of the covariant derivative $\nabla_{M}\varepsilon$ along $AdS_{2}$, $S^{5}$, and $S^{1}$ are given by,
\begin{align}
& (m) & \nabla_{m}\varepsilon & = \left( \frac{1}{f_{2}}\hat{\nabla}_{m} + \frac{D_{a}f_{2}}{2f_{2}}\Gamma_{m}\Gamma^{a} \right)\varepsilon \nonumber \\
& (i) & \nabla_{i}\varepsilon & = \left( \frac{1}{f_{5}}\hat{\nabla}_{i} + \frac{D_{a}f_{5}}{2f_{5}}\Gamma_{i}\Gamma^{a} \right)\varepsilon \nonumber \\
& (9) & \nabla_{9}\varepsilon & = \left( \frac{1}{f_{1}}\hat{\nabla}_{9} + \frac{D_{a}f_{1}}{2f_{1}}\Gamma_{9}\Gamma^{a} \right)\varepsilon
\end{align}
as well as $\nabla_{a}\varepsilon$ along $\Sigma$. The hats refer to the canonical connections on $AdS_{2}$, $S^{5}$, and $S^{1}$, respectively, and the additional term that appears in going from $\nabla$ to $\hat{\nabla}$ is due to the warp factors in the ten-dimensional metric, with $D_{a}f = \rho^{-1}\partial_{a}f$. Using the Killing spinor equations (\ref{eq:KSeqns}) to eliminate the hatted covariant derivatives, we have,
\begin{align}
& (m) & \nabla_{m}\varepsilon & = \Gamma_{m}\sum_{\eta_{1},\eta_{2},\eta_{3}}\chi^{\eta_{1},\eta_{2}}\chi^{\eta_{3}}\otimes \left( \frac{\eta_{1}}{2f_{2}}\zeta_{\eta_{1},\eta_{2},\eta_{3}} + \frac{D_{a}f_{2}}{2f_{2}}\gamma^{a}\zeta_{-\eta_{1},\eta_{2},\eta_{3}} \right)\otimes\phi^{*} \nonumber \\
& (i) & \nabla_{i}\varepsilon & = \Gamma_{i}\sum_{\eta_{1},\eta_{2},\eta_{3}}\chi^{\eta_{1},\eta_{2}}\chi^{\eta_{3}}\otimes \left( \frac{\eta_{2}}{2f_{5}}\zeta_{\eta_{1},\eta_{2},\eta_{3}} + \frac{D_{a}f_{5}}{2f_{5}}\gamma^{a}\zeta_{-\eta_{1},\eta_{2},\eta_{3}} \right)\otimes\phi^{*} \nonumber \\
& (9) & \nabla_{9}\varepsilon & = \Gamma_{9}\sum_{\eta_{1},\eta_{2},\eta_{3}}\chi^{\eta_{1},\eta_{2}}\chi^{\eta_{3}}\otimes \left( \frac{i\eta_{3}}{2f_{1}}\gamma^{9}\zeta_{-\eta_{1},\eta_{2},\eta_{3}} + \frac{D_{a}f_{1}}{2f_{1}}\gamma^{a}\zeta_{-\eta_{1},-\eta_{2},\eta_{3}} \right)\otimes\phi^{*}
\end{align}
For the additional terms involving $\varepsilon$, we project along the various directions and obtain,
\begin{align}
& (m) && \Gamma_{m}\sum_{\eta_{1},\eta_{2},\eta_{3}}\chi^{\eta_{1},\eta_{2}}\chi^{\eta_{3}}\otimes\left( \frac{1}{2}f\zeta_{\eta_{1},\eta_{2},\eta_{3}} \right)\otimes \phi^{*} \nonumber \\
& (i) && \Gamma_{i}\sum_{\eta_{1},\eta_{2},\eta_{3}}\chi^{\eta_{1},\eta_{2}}\chi^{\eta_{3}}\otimes\left( -\frac{1}{2}f\zeta_{\eta_{1},\eta_{2},\eta_{3}} \right)\otimes \phi^{*} \nonumber \\
& (a) && \sum_{\eta_{1},\eta_{2},\eta_{3}}\chi^{\eta_{1},\eta_{2}}\chi^{\eta_{3}}\otimes\left( -\frac{i}{2}q_{a} \zeta_{\eta_{1},\eta_{2},\eta_{3}} + \frac{1}{2}f\gamma^{a}\zeta_{-\eta_{1},\eta_{2},\eta_{3}} \right)\otimes \phi \nonumber \\
& (9) && \Gamma_{9}\sum_{\eta_{1},\eta_{2},\eta_{3}}\chi^{\eta_{1},\eta_{2}}\chi^{\eta_{3}}\otimes\left( \frac{1}{2}f\zeta_{\eta_{1},\eta_{2},\eta_{3}} \right)\otimes \phi^{*}
\end{align}
while for the terms involving $\varepsilon^{*}$ we have,
\begin{align}
& (m) & & \Gamma_{m}\sum_{\eta_{1},\eta_{2},\eta_{3}}\chi^{\eta_{1},\eta_{2}}\chi^{\eta_{3}}\otimes \frac{1}{16}\left( 3ig_{\bar{a}}\gamma^{\bar{a}}\star\zeta_{\eta_{1},\eta_{2},\eta_{3}} + ih\star\zeta_{-\eta_{1},\eta_{2},\eta_{3}} \right) \otimes \phi^{*} \nonumber \\
& (i) & & \Gamma_{i}\sum_{\eta_{1},\eta_{2},\eta_{3}}\chi^{\eta_{1},\eta_{2}}\chi^{\eta_{3}}\otimes \frac{1}{16}\left( -ig_{\bar{a}}\gamma^{\bar{a}}\star\zeta_{\eta_{1},\eta_{2},\eta_{3}} + ih\star\zeta_{-\eta_{1},\eta_{2},\eta_{3}} \right) \otimes \phi^{*} \nonumber \\
& (a) & & \sum_{\eta_{1},\eta_{2},\eta_{3}}\chi^{\eta_{1},\eta_{2}}\chi^{\eta_{3}}\otimes \frac{1}{16}\left[ \left( 3ig^{a} - ig_{\bar{b}}\gamma^{a\bar{b}} \right) \star\zeta_{-\eta_{1},\eta_{2},\eta_{3}} - 3ih\gamma^{a} \star\zeta_{\eta_{1},\eta_{2},\eta_{3}} \right] \otimes \phi \nonumber \\
& (9) & & \Gamma_{9}\sum_{\eta_{1},\eta_{2},\eta_{3}}\chi^{\eta_{1},\eta_{2}}\chi^{\eta_{3}}\otimes \frac{1}{16}\left[ \left( 3ig^{9}\sigma^{3} - ig_{a}\gamma^{a} \right) \star\zeta_{\eta_{1},\eta_{2},\eta_{3}} - 3ih \star\zeta_{-\eta_{1},\eta_{2},\eta_{3}} \right] \otimes \phi^{*} \nonumber \\
\end{align}
We observe that each term in the gravitino equation contains $\Gamma_{A}\chi^{\eta_{1},\eta_{2}}\chi^{\eta_{3}}$ or $\chi^{\eta_{1},\eta_{2}}\chi^{\eta_{3}}$, which we argue are linearly independent. Requiring the coefficients to vanish independently, then rewriting the relations using the $\tau$-matrix notation and eliminating the star using the definition (\ref{eq:stardef}), we recover the system of reduced gravitino BPS equations announced in (\ref{eq:gravitino}).

\bibliographystyle{JHEP.bst}
\bibliography{AdS2xS5xS1_v3}
\end{document}